\documentclass[english,prd,showpacs,amsfonts,amssymb,amsmath,twocolumn]{revtex4}
\usepackage{pslatex}
\usepackage[T1]{fontenc}
\usepackage[latin1]{inputenc}
\usepackage{babel}
\usepackage{graphics}
\usepackage{prettyref}

\makeatletter

\providecommand{\LyX}{L\kern-.1667em\lower.25em\hbox{Y}\kern-.125emX\@}
\newcommand{\noun}[1]{\textsc{#1}}

\usepackage{bm}
\usepackage{hyperref}

\makeatother
\begin{document}

\title{Optimal detection of burst events in gravitational wave interferometric
observatories}

\author{Andrea Vicer\'e}

\affiliation{LIGO Laboratory, California Institute of Technology, M/S 18-34 1200
East California Blvd, Pasadena, CA 91125, USA}

\thanks{Permanent address: Istituto di Fisica dell'Universit\'a di Urbino,
via S.Chiara 27, I-61029 Urbino, Italy}

\email{vicere@fis.uniurb.it}

\preprint{LIGO-P010019-01-E}

\date{\today}

\begin{abstract}
We consider the problem of detecting a burst signal of unknown shape
in the data from gravitational wave interferometric detectors. We
introduce a statistic which generalizes the \emph{excess power} statistic
proposed first by Flanagan and Hughes, and then extended by Anderson
\emph{et al.} also to a multiple detector case. The statistic that
we propose is shown to be optimal for arbitrary noise spectral characteristic,
under the two hypotheses that the noise is Gaussian, albeit colored,
and that the prior for the signal is uniform.

The statistic derivation is based on the assumption that a signal
affects only affects \( N_{\parallel } \) samples in the data stream,
but that no other information is a priori available, and that the
value of the signal at each sample can be arbitrary. This is the main
difference from previous works, where different assumptions were made,
like a signal distribution uniform with respect to the metric induced
by the (inverse) noise correlation matrix. The two choices are equivalent
if the noise is \emph{white}, and in that limit the two statistics
do indeed coincide. In the general case, we believe that the statistic
we propose may be more appropriate, because it does not reflect the
characteristics of the noise affecting the detector on the supposed
distribution of the gravitational wave signal.

Moreover we show that the proposed statistic can be easily implemented
in its exact form, combining standard time-series analysis tools which
can be efficiently implemented, and the resulting computational cost
is still compatible with an \emph{on-line} analysis of interferometric
data.

We generalize this version of an excess power statistic to the multiple
detector case, considering first the noise uncorrelated among the
different instruments, and then including the effect of correlated
noise: we show that this can be done either perturbatively, or in
exact form.

We give full details about the implementation of the algorithm, both
for the single and the multiple detector case, and we discuss exact
and approximate forms; the choice among them depends on the specific
characteristics of the noise and on the assumed length of the burst
event.

As a example, we show what would be the sensitivity of the network
of interferometers to a \( \delta  \)-function burst.
\end{abstract}


\maketitle

\section{Introduction and summary}

Several large scale interferometric detectors \cite{LIGO:92,VIRGO:97,GEO:95,TAMA}
are currently under commissioning and are expected to start data acquisition
and reach their design sensitivity in a few years. Some of the candidate
sources, like the coalescing binaries in their inspiral phase, can
be modeled with reasonable accuracy and the gravitational waveforms
can be predicted, thus allowing a matched-filter detection strategy;
see \cite{owen:sathya:99} and references therein for a review. On
the other hand, as argued in \cite{pradier:01} it is conceivable
that the uncertainty on the waveform will remain high for sources
like the Type II supernova explosions or the merger phase in the coalescence
of black holes or neutron stars: in this context, the issue of detecting
events poorly modeled or not modeled at all remains crucial.

The problem has already been faced from different point of views:
some authors \cite{pradier:00,pradier:01,arnaud:01} aim at devising
several simple and computational inexpensive algorithms, to be run
in parallel after having been tested and optimized against model waveforms
\cite{zwerger:mueller:93}. Others start from general hypotheses on
the distribution of the signals and the noise and derive statistics
optimal under those assumptions \cite{flanagan:hughes:98,arnaud:99,anderson:brady:creighton:flanagan:00,anderson:brady:creighton:flanagan:01}.

We consider of particular interest strategies, like the \emph{excess
power} statistic proposed by Flanagan and Hughes \cite{flanagan:hughes:98},
or the \emph{norm filter} studied by Arnaud \emph{et al.} \cite{arnaud:99},
which try to make minimal assumptions on the nature of the signal,
like time duration and bandwidth only: in particular the excess power
statistic has been recently analyzed by Anderson \emph{et al.} \cite{anderson:brady:creighton:flanagan:01}
and extended to the {}``blind'' search of burst events from a network
of interferometers. However, as pointed out in \cite[note 8]{anderson:brady:creighton:flanagan:01},
the authors have actually made the assumption that the signal distribution
is flat with respect to the inner product defined by the inverse of
the noise correlation matrix. They recognize that this is an approximation,
and correctly claim that it is legitimate when the noise spectrum
does not vary rapidly in the band of interest: however we will argue
in \prettyref{sec:generalities} that this may be not true for real
detector noise, and we will show that in view of the current models
for burst signals from the core collapse of supernovae \cite{zwerger:mueller:93,dimmelmeier:font:mueller:01}
the correlation length of the detector noise cannot be assumed short
with respect to the event duration, even assuming the design noise.

The choice of the signal distribution in \cite{anderson:brady:creighton:flanagan:01}
has the advantage of making easy to incorporate a priori informations,
when available, about the absolute scale of the expected signals:
we shall see that this is in general more complicated with our statistic,
if detection thresholds were to be set using a Bayesian criterion.

The analysis method that we propose consists essentially of two steps:

\begin{description}
\item [A]filter the input data with a matched filter for \( \delta  \)
functions.
\item [B]Compute a statistic similar to the \emph{energy} of the data within
each time window we are willing to test for the presence of a burst,
using a particular scalar product which can be conveniently computed
either using the discrete Karhunen Lo\`eve transform (DKLT), or (approximately)
using a discrete Fourier transform (DFT).
\end{description}
The algorithm can be easily generalized to the multiple detector case,
resulting in a optimal statistic which depends on the direction in
the sky the signal is supposed to come from. It also turns out that
possible correlations among the detector noises can be taken into
account in a natural way by modifying the \( \delta  \)-filtering
step, and that these modifications are a natural consequence of the
likelihood maximization procedure.

The paper is organized in two main sections: in \prettyref{sec:singleDetector}
we derive the statistic for the case of a single detector: then in
\prettyref{sec:multipleDetector} we extend it to the multiple detector
case.

In greater detail, the plan of the work is as follows: in \prettyref{sec:generalities}
we motivate our study, introducing our hypotheses on the signal and
the noise and discussing them in view of the current models for supernovae
signals. In \prettyref{sec:detectionFramework} we briefly recall
the Bayesian framework we follow in deriving the optimal statistic.
In \prettyref{sec:burstUnit} we consider the simple but illuminating
case in which the burst event is a \( \delta  \) function, that is
with a duration affecting only one data sample: we deduce a detection
strategy that corresponds to one of the data analysis methods used
in resonant bar experiments \cite{pizzella:93}. In \prettyref{sec:burstIntegration}
we derive the exact expression for the likelihood ratio, for a burst
affecting \( N_{\parallel } \) samples in the data stream, and we
show that the calculational steps are essentially the two \textbf{A,B}
mentioned above. In \prettyref{sec:KLexpansion} we make a digression
on the Karhunen-Lo\`eve transform, a tool well known in statistics
\cite{papoulis:91,therrien:92} and already applied in the analysis
of data from bar detectors \cite{ortolan:vedovato:cerdonio:vitale:94},
and proposed for the study of narrow resonances \cite{cella:chickarmane:vicere:00}:
in \prettyref{sec:exactStatistic} we apply the DKLT expansion to
our problem and we show that it is a convenient way to implement step
\textbf{B}. When the supposed burst length \( N_{\parallel } \) is
large, an approximate formula using the DFT can be used, as discussed
in \prettyref{sec:approximateStatistic}.

The very meaning of the word {}``optimal'' used for defining the
proposed statistic is discussed in~\prettyref{sec:discussion}, where
we clarify the limits of the method, in particular with respect to
the possible inclusion of prior information on the strength of the
signal.

We elaborate in \prettyref{sec:implementation} on the details of
the detection algorithm, and we show that step \textbf{A} is easily
implemented using standard tools of time series analysis, including
the whitening transformation, although the required operation is different
from a whitening.

The distribution of our statistic in absence and presence of a signal
is considered in~\prettyref{sec:statisticalAnalysis}: we show that
it corresponds exactly to a \( \chi ^{2} \) variable, respectively
central or not central.

The multiple detector case is treated first in the approximation of
uncorrelated Gaussian noise across the detector network in \prettyref{sec:networkUncorrelatedNoise},
where we show that the techniques used in the single detector case
can be easily extended, in a way not much different from what has
been done in \cite{anderson:brady:creighton:flanagan:01}, but including
the above mentioned step \textbf{A} implementing the filtering for
\( \delta  \) functions. We show in \prettyref{sec:generalNetworkCase}
that the step \textbf{B} of \textbf{}the \textbf{}resulting algorithm
can be written in exact form using a vector DKLT, while a simpler
form, similar to the one derived in \cite{anderson:brady:creighton:flanagan:01},
is valid in the long burst limit, as discussed in \prettyref{sec:simplifiedNetworkCase}.

The case in which the noise of different detectors displays some degree
of correlation is considered in~\prettyref{sec:networkCorrelatedNoise}:
we start from the same hypotheses as Finn \cite{finn:01} and we show
how the effect of the cross-detector terms in the noise correlation
matrix can be taken into account in our algorithm, either perturbatively,
if the cross terms are small, or exactly if they are not: not surprisingly,
the needed modifications turn out to affect only the \( \delta  \)-filtering
step, and in a simple way.

We finally give an example of application of the algorithm to the
detection of bursts of unit duration, for a network of detectors comprising
either the three LIGOs \cite{LIGO:92} or including also GEO600 \cite{GEO:95},
TAMA \cite{TAMA} and Virgo \cite{VIRGO:97}. We compute the resulting
\textsf{SNR}, which is a function of the direction in the sky; this
allows us to pictorially show to which extent the network analysis
strategy is advantageous, at least in the ideal situation in which
the detector noise is Gaussian.

Throughout all the paper we adopt a discrete-time, discrete-frequency
notation: the conventions followed for the Discrete Fourier Transform
require some care and are detailed in App.~\ref{sec:conventions},
while the characteristics of the detectors in the network are detailed
in App.~\ref{sec:networkModel}.

\section{\label{sec:singleDetector}Single detector analysis}

\subsection{\label{sec:generalities}Noise and signal statistics}

We will keep consistently a discrete time, discrete frequency notation,
assuming a sampling rate \( f_{s} \) and a finite observation time
\( T=N/f_{s} \). We assume that the detector noise has zero mean
and is Gaussian, albeit colored, characterized by a correlation matrix\begin{equation}
\left( \mathbf{R}_{n}\right) \left[ i,j\right] =E[n_{i}n_{j}]
\end{equation}
where \( n_{i}\equiv n[i]\equiv n\left( i/f_{s}\right)  \), and \( i\in [1,N] \).
We further assume that the noise is stationary, hence \( \mathbf{R} \)
is a symmetric Toeplitz matrix \cite{therrien:92}, whose entries
depend only on the difference of indices. In terms of this correlation
matrix, the probability of observing a certain set of noise data \( \mathbf{n} \)
(of total length \( N) \) is given by the joint distribution

\begin{equation}
\label{eq:Pnoise}
P(\mathbf{n})=\frac{1}{\sqrt{(2\pi )^{N}\det R}}\exp \left[ -\frac{1}{2}n_{i}(\mathbf{R}_{n}^{-1})_{i\, j}n_{j}\right] ;
\end{equation}
if a certain signal \( \mathbf{s} \) is also present, the conditioned
probability of observing a set of data \( \mathbf{x} \) is\begin{equation}
\label{eq:PnoiseSignal}
P(\mathbf{x}|\mathbf{s})=\frac{1}{\sqrt{(2\pi )^{N}\det R}}e^{\left[ -\frac{1}{2}(\mathbf{x}-\mathbf{s})\cdot \mathbf{R}_{n}^{-1}\cdot (\mathbf{x}-\mathbf{s})\right] }\, .
\end{equation}
We should stress that formulas in Eqs.~(\ref{eq:Pnoise},\ref{eq:PnoiseSignal})
make use of the information available in the finite data sequence:
we observe only \( N \) data and with the expression of \( P(\mathbf{n}) \)
and \( P(\mathbf{x}|\mathbf{s}) \) we cannot take into account the
effect of the past data points, which fall outside our observation
window%
\footnote{We are assuming that the noise processes are causal: real noise is
expected to be such, unless it is processed with a non-causal filter.
For instance a matched filter is in general non-causal, because its
impulse response is non-zero for negative times\cite[sec. 6.5]{donough:whalen:95}.
}. We shall see later that when considering a shorter analysis window
(of size \( N_{\parallel } \)) contained in a longer data train we
should actually exploit also the information contained outside the
\( N_{\parallel } \) window. This information is of little relevance
if the analysis window is much longer than the largest correlation
times in the noise: but this is not generally the case when considering
burst events.

What matters in deciding what is the relevance of this boundary effect
is the noise spectrum: one knows indeed (see \prettyref{sec:conventions})
that \begin{equation}
\label{eq:inverseMatrixR}
R^{-1}\left( \frac{a-b}{f_{s}}\right) \equiv \left( \mathbf{R}_{n}^{-1}\right) _{ab}=\frac{2}{f_{s}N}\sum ^{N-2}_{k=1}\frac{e^{i2\pi k\left( a-b\right) }}{S_{n}\left[ k\right] }\, ,
\end{equation}
and considering a noise model as given in Eq.~\prettyref{eq:noiseModel},
which summarizes the best sensitivity reachable in the first generation
interferometers, one deduces that

\begin{equation}
R^{-1}\left( \tau \right) \simeq A_{0}e^{-\left| t\right| /\tau _{0}}\cos 2\pi f_{0}\tau +\ldots 
\end{equation}
 neglecting faster decaying terms. The decay time(s) \( \tau _{0,1,\ldots } \)
characterize how much the matrix \( \mathbf{R}^{-1} \) (and therefore
\( \mathbf{R} \) itself) differs from a diagonal matrix. In \prettyref{sec:networkCharacteristics}
and in Table~\ref{tab:ringingParameters} we show that for the current
models of the baseline interferometer noise in the first generation
detectors the values of \( \tau _{0} \) range from 1.4 ms in TAMA
to 6ms in Virgo (corresponding to O(100) samples).

These time scales should be compared with the expected duration of
the bursts: for instance Zwerger and Müller \cite[Figs. 5,6]{zwerger:mueller:93}
have shown several examples of gravitational waveforms emitted in
axisymmetric core collapse events, displaying large variations on
scales of a few ms, and narrow large amplitude peaks even shorter
than 1~ms. The same features are found in more recent simulations,
which include relativistic effects, by Dimmelmeier \emph{et al}.\cite[Fig. 2 (Model A)]{dimmelmeier:font:mueller:01}:
we conclude that it is generally not justified from the physical point
of view to surmise that the correlation decay times of the noise are
short compared to the burst duration.

Moreover, the values for the decay times quoted in Table~\ref{tab:ringingParameters}
refer to a ideal detector noise, free of narrow resonances: real detectors
may well exhibit richer spectral features \cite{ligo40mData:96}.
For instance a thermal resonance with proper frequency \( f_{0} \)
and quality factor \( Q \) would contribute to the noise correlation
a term with a characteristic decay time \( \tau _{D}=\frac{Q}{\pi f_{0}} \):
with violin modes easily having \( Q>10^{5} \), and frequencies \( f_{0}=O(1000) \)
Hz, the decay time can easily reach tens of seconds. Although it is
foreseen to subtract the effect of these resonances from the data,
using for instance Kalman filters \cite{finn:mukherjee:01} it is
fair to say that any residual effect, due for instance to a imperfect
cancellation of a very high \( Q \) resonance, will contribute to
increase the noise correlation length above the values deduced from
the baseline noise.

There are at least two important consequences of the presence of a
non-zero correlation length: (1) statistics built using maximum likelihood
criteria must be modified to take into account noise outside the window
affected by the burst event; (2) instances of these statistics, derived
from these data, will exhibit a correlation in time which will have
to be kept into account when computing false alarm and false dismissal
probabilities \cite{schutz:91}. In the present paper we will concentrate
on the first issue.

One general way, other than subtracting the narrow spectral components,
to attack this problem would be to assume that data have been pre-whitened
\cite{arnaud:01,arnaud:99,pradier:00,pradier:01}, for instance using
a time-domain filter estimated from the data themselves \cite{cuoco:01,cuoco:02}:
however this strategy requires to take into account the effect of
the whitening filter throughout the whole detection chain, in particular
the alteration of the signal waveform and consequently of the signal
distribution. In other words, when integrating over the space of possible
signals, we are not allowed to ignore that the measure is changed
by the transformation.

To render our statements more precise, we need to discuss in detail
the detection framework adopted.

\subsection{\label{sec:detectionFramework}Detection framework}

We suppose that, when present, the burst affects only a interval \( T_{\parallel }=N_{\parallel }/f_{s} \)
starting at absolute time \( t_{\mathrm{burst}} \), that is a number
\( N_{\parallel } \) of samples in the data stream. We are unable
to prescribe any prior for the signal amplitudes, so we treat it,
at any given instant, as \emph{nuisance} parameters, to be integrated
out. Anderson \emph{et al.} \cite{anderson:brady:creighton:flanagan:01,flanagan:hughes:98}
have made similar assumptions, but as we have already anticipated
with a different hypothesis on the measure for this integration.

Given a data vector \( \mathbf{x} \), the \emph{a posteriori} probability
of having observed those values can be written as\cite{finn:92}

\begin{equation}
P(\mathbf{x})=P(\mathbf{x}|1)P(1)+P(\mathbf{x}|0)P(0)
\end{equation}
where as usual \( P(1),P(0) \) are the \emph{a priori} probabilities
for a signal of unspecified form being present in the data, and \( P(\mathbf{x}|1) \)
is the probability of observing the dataset \textbf{\( \mathbf{x} \)}
given that \emph{some} signal \( \mathbf{s} \) is present, while
\( P(\mathbf{x}|0) \) is the probability of observing the same dataset
\( \mathbf{x} \) in absence of signals: it was defined in Eq.~\prettyref{eq:Pnoise}.
In turn we have that\begin{equation}
P(\mathbf{x}|1)\equiv \int P(\mathbf{x}|\mathbf{s})P(\mathbf{s})\, d\mathbf{s}
\end{equation}
where \( P(\mathbf{s}) \) is the \emph{a-priori} probability of having
a signal \( \mathbf{s} \) present, while \( P(\mathbf{x}|\mathbf{s}) \)
has already been defined in Eq.~\prettyref{eq:PnoiseSignal}. We
may or may not have a good guess for the distribution \( P(\mathbf{s}) \):
in this paper we assume to have no \emph{a priori} information.

Given our complete ignorance of \textbf{\( P(1),P(0) \)}, we resort
to defining the (integrated) likelihood ratio that \emph{any} signal
is present as\noun{\begin{equation}
\label{eq:LikelyhoodRatio}
\Lambda (\mathbf{x})\equiv \frac{P(\mathbf{x}|1)}{P(\mathbf{x}|0)}=\int e^{-\frac{1}{2}\mathbf{s}\cdot \mathbf{R}_{n}^{-1}\cdot \mathbf{s}+\mathbf{s}\cdot \mathbf{R}_{n}^{-1}\cdot \mathbf{x}}P(\mathbf{s})d\mathbf{s};
\end{equation}
}in terms of \( \Lambda  \) and using the Bayes rule \( P(1|\mathbf{x})P(\mathbf{x})=P(\mathbf{x}|1)P(1) \)
we can write\begin{equation}
P(1|\mathbf{x})=\frac{\Lambda \left( \mathbf{x}\right) }{\Lambda \left( \mathbf{x}\right) +P(0)/P(1)}\, ;
\end{equation}
 for the probability of having observed a signal, conditioned by the
particular instance of data \( \mathbf{x} \) that we have received.
Although we have no idea of the priors \( P(0),P(1) \), this probability
is a monotonic function of the likelihood \( \Lambda \left( \mathbf{x}\right)  \),
which is therefore the quantity to be estimated, in dependence on
the assumptions (or lack of assumptions) on \( P\left( \mathbf{s}\right)  \):
to implement our complete ignorance on the waveform we will assume
that \( P\left( \mathbf{s}\right)  \) is flat in the space \( \mathbb R^{N_{\parallel }} \)
of possible signals of length \( N_{\parallel } \).

For the sake of comparison, Anderson \emph{et al.} \cite[Eq. 3.1, and the following discussion and note]{anderson:brady:creighton:flanagan:01}
have assumed that signals \( \mathbf{s} \) are a priori distributed
in a uniform way with respect to the metric induced by the scalar
product \( \left\langle x,y\right\rangle \equiv \mathbf{x}\cdot \mathbf{R}^{-1}_{n}\cdot \mathbf{y} \):
this is the essential reason why our results are different from theirs.

\subsection{\label{sec:burstUnit}Burst of unit duration (\protect\( \delta \protect \)
function)}

Let us first consider a extreme example, namely a burst affecting
only one of our data samples: its amplitude is unknown, and we want
to test the hypothesis that the event occurs at the sample arbitrarily
labeled \( a \). Then the integrated likelihood ratio in Eq.~\prettyref{eq:LikelyhoodRatio}
is \begin{eqnarray}
\Lambda (\mathbf{x}) & = & \int ^{+\infty }_{-\infty }e^{-\frac{1}{2}s_{a}(\mathbf{R}_{n}^{-1})_{aa}s_{a}+s_{a}(\mathbf{R}_{n}^{-1}\cdot \mathbf{x})_{a}}\, ds_{a}\nonumber \\
 & = & \exp \left[ \frac{1}{2}\frac{\left[ (\mathbf{R}_{n}^{-1}\cdot \mathbf{x})_{a}\right] ^{2}}{(\mathbf{R}_{n}^{-1})_{aa}}\right] \, ,
\end{eqnarray}
 where no summation over \( a \) is understood. The interesting statistic
for a certain possible arrival index \( a \) is \begin{equation}
\label{eq:WienerDelta}
L_{a}\left( \mathbf{x}\right) \equiv 2\ln \Lambda \left( \mathbf{x}\right) =\frac{f_{s}}{N}\frac{\left| \sum _{k}e^{i2\pi ak/N}\, \frac{\tilde{x}\left[ k\right] }{S_{n}\left[ k\right] }\right| ^{2}}{\sum _{k}\frac{1}{S_{n}\left[ k\right] }},
\end{equation}
and is readily identified as the Wiener filter for a \( \delta  \)
function\cite{pizzella:93}. In absence of signals its distribution
is \begin{equation}
\label{eq:distributionChi2}
d_{0}(L)=\frac{1}{\sqrt{2\pi \, L}}e^{-\frac{1}{2}L}
\end{equation}
as expected for the square of a Gaussian variable; note that the statistic
\( L \) is \emph{not} equivalent to taking the energy of the signal. 

Nor is it equivalent to first whitening the data and then taking the
energy: indeed whitening would mean, in matrix notation, to perform
a lower-upper (LU) factorization of the matrix \( \mathbf{R}^{-1} \)
in the form%
\footnote{It is well known\cite[chap. 2]{therrien:92} that a correlation matrix
$\mathbf{R}$, being hermitian and symmetric, can be expressed as $\mathbf{R} = \mathbf{L}\cdot\mathbf{D}_L\cdot\mathbf{L}^t$,
where $\mathbf{L}$ is a unit lower matrix and $\mathbf{D}_L$ is a
diagonal matrix. This implies that $\mathbf{R}^{-1}=(\mathbf{L}^t)^{-1}\cdot\mathbf{D}^{-1}_L\cdot\mathbf{L}^{-1}$,
and note that also $\mathbf{L}^{-1}$ is unit lower triangular. For
our purposes we can take the square root of the diagonal matrix $\mathbf{D}_L$
so as to define $\mathbf{W}\equiv\sqrt{\mathbf{D}_L^{-1}}\cdot\mathbf{L}^{-1}$
and obtain $\mathbf{R}^{-1}=\mathbf{W}^t\cdot\mathbf{W}$.
}\begin{equation}
\label{eq:spectralFactorization}
\mathbf{R}_{n}^{-1}=\mathbf{W}_{n}^{t}\cdot \mathbf{W}_{n}\, ;
\end{equation}
where \( \mathbf{W}_{n} \) is a lower triangular matrix which defines
a causal transformation, dependent on the noise statistics: the matrix
\( \mathbf{W}_{n} \) defines a whitening transformation, namely the
random vector \( \mathbf{z}=\mathbf{W}_{n}\cdot \mathbf{x} \) is
distributed as white gaussian noise. The statistic \( L_{a}\left( \mathbf{x}\right)  \)
can also be written in terms of the whitened data \( \mathbf{z} \)

\begin{equation}
L_{a}(\mathbf{x})=\frac{\left[ \left( \mathbf{W}_{n}^{t}\cdot \mathbf{z}\right) _{a}\right] ^{2}}{(\mathbf{W}_{n}^{t}\cdot \mathbf{W}_{n})_{aa}}\, ,
\end{equation}
which can be regarded as a en\emph{ergy}, but takes into account the
altered signal shape (a \( \delta  \) function in this case) under
the application of the transformation \( \mathbf{W}_{n} \).

\emph{In passim} it is worth noticing that we wrote the distribution
\( d_{0}(L) \) assuming implicitly to compute the moments of our
statistic over a ensemble of noise realizations; in practice however
one takes samples from a single time-series, at different locations
in the data stream. This means that successive instances of the statistic
\( L, \) for different indices \( a \), will belong to a joint distribution,
which does not factor in a product of terms like the \( \chi ^{2} \)
in Eq.~\ref{eq:distributionChi2}: for instance we can consider the
cumulant\begin{equation}
\frac{E\left[ L_{a}\left( \mathbf{x}\right) L_{b}\left( \mathbf{x}\right) \right] }{E\left[ L_{a}\left( \mathbf{x}\right) \right] E\left[ L_{b}\left( \mathbf{x}\right) \right] }-1=2\left[ \frac{\left( \mathbf{R}_{n}^{-1}\right) _{ab}}{\left( \mathbf{R}_{n}^{-1}\right) _{aa}}\right] ^{2},
\end{equation}
which exposes that the statistic \( L_{a} \) as a function of the
supposed burst event location \( a \) has a non-negligible cross
correlation in colored noise. This should be always kept in mind when
post-processing the results of the analysis, for instance when deriving
limits on false alarm or false dismissal rates\cite{schutz:91}: we
will not elaborate further on this.

We now proceed to generalize the statistic \( L \) to a burst of
arbitrary length.

\subsection{\label{sec:burstIntegration}General burst}

It is useful to introduce a vector space notation: let us call \( {\cal V}_{N} \)
the vector space of all possible data vectors, having length \( N \);
we are willing to test for the hypothesis of presence of a burst signal
in a certain subspace \( {\cal V}_{\parallel } \), defined by taking
\( N_{\parallel }\ll N \) consecutive samples, from a certain starting
position (say, \( l \)) in the original vector. The orthogonal subspace,
of dimension \( N-N_{\parallel }\sim N \), will be called \( {\cal V}_{\perp } \).

We have again the likelihood ratio\begin{equation}
\Lambda (\mathbf{x})=\int e^{-\frac{1}{2}s_{i}(\mathbf{R}_{n}^{-1})_{ij}s_{j}+s_{i}(\mathbf{R}_{n}^{-1}\cdot \mathbf{x})_{i}}\prod ^{N_{\parallel }}_{i}ds_{i}
\end{equation}
where the indices \( i,j \) run only over elements of \( {\cal V}_{\parallel } \);
however the noise correlation matrix \( \mathbf{R}_{n} \) is defined
for an arbitrary index difference. Let us introduce the matrix\textbf{\begin{equation}
\label{eq:defM}
(\mathbf{R}_{n}^{-1})_{\parallel }\equiv (\mathbf{R}_{n}^{-1})_{[l:l+N_{burst},\, l:l+N_{burst}]}
\end{equation}
}obtained restricting the indices of the \( \mathbf{R}_{n}^{-1} \)
matrix on the \textbf{\( [l,l+N_{burst}] \)} interval: it acts only
on the \( {\cal V}_{\parallel } \) subspace.

\begin{widetext}Performing the Gaussian integrals over amplitudes
\( s_{i} \) of the signal, which we treat as \emph{nuisance} parameters,
we obtain\begin{equation}
\label{eq:Lambda}
\Lambda \left( \mathbf{x}\right) \propto \exp \left[ \frac{1}{2}(\mathbf{R}_{n}^{-1}\cdot \mathbf{x})_{\parallel }\cdot \left( \left( \mathbf{R}_{n}^{-1}\right) _{\parallel }\right) ^{-1}\cdot (\mathbf{R}_{n}^{-1}\cdot \mathbf{x})_{\parallel }\right] =\exp \left[ \frac{1}{2}x_{\alpha }(\mathbf{R}_{n}^{-1})_{\alpha i}\left( \left( \left( \mathbf{R}_{n}^{-1}\right) _{\parallel }\right) ^{-1}\right) _{ij}(\mathbf{R}_{n}^{-1})_{j\beta }x_{\beta }\right] 
\end{equation}
where indices \( \alpha ,\beta  \) run in \( {\cal V}_{\parallel }+{\cal V}_{\perp } \),
and indices \( i,j \) run in \( {\cal V}_{\parallel } \); the overall
normalization does not depend on \( \mathbf{x} \) and can be ignored.

\end{widetext}We are forced to this somewhat cumbersome notation because
the operations of projecting over the {}``burst'' subspace \( {\cal V}_{\parallel } \)
and of inverting a matrix do not commute. They do only when \( N=N_{\parallel } \):
in that case one would have\begin{equation}
\Lambda (\mathbf{x})\propto e^{\frac{1}{2}\mathbf{x}\cdot \mathbf{R}_{n}^{-1}\cdot \mathbf{x}}=e^{\frac{1}{2}\left( \mathbf{W}_{n}\cdot \mathbf{x}\right) ^{2}}
\end{equation}
where we used Eq.~\prettyref{eq:spectralFactorization}, and \( \mathbf{W}_{n}\cdot \mathbf{x} \)
is distributed as white gaussian noise. This means that in the case
\( N=N_{\parallel } \) the likelihood would be just a monotonic function
of the \emph{energy} \( (\mathbf{W}_{n}\cdot \mathbf{x})^{2} \) of
the whitened sample. Note however that this example is very different
from the previous one: there we had a unit burst in a long data train,
here we would have a burst as long as the data train: both are extreme
cases.

We will from now on instead take as optimal statistic for a unknown
burst, with flat prior for the sample amplitudes, the expression\begin{equation}
\label{eq:exactStatistic}
L(\mathbf{x})\equiv \sum _{i,j\in {\cal V}_{\parallel }}(\mathbf{R}_{n}^{-1}\mathbf{x})_{i}\left( \left( \left( \mathbf{R}_{n}^{-1}\right) _{\parallel }\right) ^{-1}\right) _{ij}(\mathbf{R}_{n}^{-1}\mathbf{x})_{j}\, .
\end{equation}
The reader might wonder if this is of any practical use. In particular,
the estimation of the matrix \( \left( \left( \mathbf{R}_{n}^{-1}\right) _{\parallel }\right) ^{-1} \)
looks awkward: we should estimate \textbf{\( \mathbf{R}_{n} \)},
then compute \( \mathbf{R}_{n}^{-1} \), then take a minor \( N_{\parallel }\times N_{\parallel } \)
along the matrix diagonal and finally invert it%
\footnote{It is also worth noting that while $\mathbf{R}_n$ is a Toeplitz matrix, $\mathbf{R}^{-1}_n$
is not necessarily so, though it converges to a Toeplitz matrix in
the $N\rightarrow\infty$ limit.
}.

Let us however notice the trivial identity\begin{eqnarray}
E\left[ \left( \mathbf{R}_{n}^{-1}\mathbf{n}\right) _{i}\left( \mathbf{R}_{n}^{-1}\mathbf{n}\right) _{j}\right]  & = & \sum _{\alpha \, \beta }\left( \mathbf{R}_{n}^{-1}\right) _{i\alpha }\left( \mathbf{R}_{n}^{-1}\right) _{j\beta }E\left[ n_{\alpha }n_{\beta }\right] \nonumber \\
 & = & \left( \mathbf{R}_{n}^{-1}\right) _{ij}\label{eq:RinverseFromRmx} 
\end{eqnarray}
in other words, considering the time series \( \mathbf{R}_{n}^{-1}\cdot \mathbf{n} \),
which is \( \mathbf{R}_{n}^{-1}\cdot \mathbf{x} \) in absence of
signal, and restricting it to the \( {\cal V}_{\parallel } \) subspace,
we obtain an autocorrelation matrix that is just \( \mathbf{R}_{n}^{-1} \).
Barring boundary effects on the matrix \( \mathbf{R}_{n} \), also
the process \( \mathbf{R}_{n}^{-1}\cdot \mathbf{n} \) is shift invariant,
and \( \mathbf{R}_{n}^{-1} \) is a Toeplitz matrix when considering
diagonal minors sufficiently far from the borders. 

We are therefore able to easily compute \( \mathbf{R}_{n}^{-1} \)
just from the analysis of the series \( \mathbf{R}_{n}^{-1}\cdot \mathbf{n} \):
the last step, in order to write down the statistic for a burst in
a certain subspace \( {\cal V}_{\parallel } \), is to restrict the
matrix to that interval and invert it: a efficient tool to accomplish
this task is the Karhunen-Lo\`eve expansion\cite{therrien:92}, which
we find useful to recall briefly in the next section.

\subsection{\label{sec:KLexpansion}Karhunen-Lo\`eve expansion}

In \prettyref{sec:burstUnit} we have exploited the LU factorization
of a correlation matrix \( \mathbf{R} \). It is also well known \cite{donough:whalen:95,papoulis:91,therrien:92}
that, being \( \mathbf{R} \) real symmetric and positive definite,
an expansion in terms of its eigenvalues and eigenvectors exists,
namely\begin{equation}
\label{eq:RfullDKL}
R_{\alpha \beta }=\sum ^{K}_{k=1}\sigma _{k}\psi ^{k}_{\alpha }\psi ^{k}_{\beta }
\end{equation}
where \( K \) is the dimension of the matrix and\begin{equation}
R_{\alpha \beta }\psi _{\beta }^{k}=\sigma _{k}\psi ^{k}_{\beta }
\end{equation}
with eigenvalues \( \sigma _{k}>0 \). The \( \left\{ \bm \psi ^{k},\, k\in \left[ 1,K\right] \right\}  \)
eigen-vectors are chosen orthonormal\begin{equation}
\sum _{\alpha }\psi ^{k}_{\alpha }\psi ^{l}_{\alpha }=\delta ^{kl};
\end{equation}
and define a basis in the space \( \mathbb R^{K} \): any data vector
\( \mathbf{x} \) can be written as\begin{equation}
\mathbf{x}=\sum ^{K-1}_{k=0}c_{k}\bm \psi ^{k},\, \, \mathrm{where}\, \, c_{k}\equiv \mathbf{x}\cdot \bm \psi ^{k}\, ;
\end{equation}
this decomposition is called discrete Karhunen-Lo\`eve transform (DKLT).
As with the Fourier transform, the Parseval's theorem holds\begin{equation}
\mathbf{x}\cdot \mathbf{x}=\sum ^{K}_{k=1}c^{2}_{k}
\end{equation}
and it is immediate to show that\begin{subequations}\begin{eqnarray}
E[c_{k}\, c_{l}] & = & \sigma _{k}\delta _{kl}\label{eq:avDKLTcoeff} \\
E\left[ \mathbf{x}\cdot \mathbf{x}\right]  & = & \sum ^{K}_{\alpha =1}R_{\alpha \alpha }=\sum ^{K}_{k=1}\sigma _{k}\, .\label{eq:poissonDKLT} 
\end{eqnarray}
\end{subequations}The similarity with the Fourier transform goes further:
it can be shown (see for instance \cite[sec. 4.7.2]{therrien:92})
that in the limit of large \( K \) the basis elements converge to
sines and cosines, and the eigenvalues converge to the corresponding
bins of the spectral density.

However, for finite \( K \) the DKLT is a better representation for
the noise because it takes into account the finite-size effects \cite{ortolan:vedovato:cerdonio:vitale:94}:
recall that we are interested in \( N_{\parallel } \) not necessarily
large. In particular, the coefficients \( c_{k} \) are uncorrelated
random variables, thus making the statistical analysis easier.

We are naturally led to apply the DKLT to the problem at hand.

\subsection{\label{sec:exactStatistic}Exact expression for the burst statistic}

In \prettyref{sec:burstIntegration} we have shown that the exact
statistic, defined in Eq.~\prettyref{eq:exactStatistic}, can be
expressed a\begin{equation}
\label{eq:statisticWithInverse}
L=\mathbf{y}_{\parallel }\cdot \left[ \left( \mathbf{R}_{y}\right) _{\parallel }\right] ^{-1}\cdot \mathbf{y}_{\parallel }
\end{equation}
where \( \mathbf{y}_{\parallel }\equiv \left( \mathbf{R}_{n}^{-1}\cdot \mathbf{x}\right) _{\parallel } \).

The vector \( \mathbf{y} \) itself can be easily computed: recall
that we have assumed \( N \) large, hence we can write\begin{equation}
y[l]=\frac{1}{N}\sum ^{N-1}_{k=0}\frac{2}{S_{n}\left[ k\right] }e^{-i2\pi \, k\, l/N}\tilde{x}[k]\, ;
\end{equation}
where \( S_{n} \) is the one-sided spectrum corresponding to the
correlation matrix \( \mathbf{R}_{n} \).

Thanks to the KL transform we are now also able to write down explicitly
the inverse of the correlation matrix \( \mathbf{R}_{y} \) of the
\( \mathbf{y} \) time series, keeping into account the restriction
to the \( {\cal V}_{\parallel } \) space, in terms of a appropriate
DKL basis \( {\bm \psi _{\parallel }^{k},\, k\in \left[ 1,\, N_{\parallel }\right] } \),
as\begin{equation}
\left[ \left( \mathbf{R}_{y}\right) _{\parallel }\right] ^{-1}=\sum ^{N_{\parallel }}_{k=1}\frac{1}{\sigma _{k}}\bm \psi _{\parallel }^{k}\otimes \bm \psi _{\parallel }^{k};
\end{equation}
hence we finally have \begin{equation}
\label{eq:statisticDKLT}
L=\sum ^{N_{\parallel }}_{k=1}\frac{1}{\sigma _{k}}\left( \bm \psi ^{k}_{\parallel }\cdot \mathbf{y}_{\parallel }\right) ^{2}.
\end{equation}
Note that the matrix \( \mathbf{R}_{y} \), as we have seen in \prettyref{sec:burstIntegration},
is approximately a Toeplitz matrix: this means that \( \left( \mathbf{R}_{y}\right) _{\parallel } \)
does not depend on the segment, in the data train, chosen to test
for the presence of bursts%
\footnote{This approximation will be better far from the boundaries of the data
train, a fact which suggests, in the analysis of a continuous stream
of data, to work on (long) overlapping segments; another consequence
is that the DKL basis can be estimated once and repeatedly used for
subsequent data segments.
}. 

The reader might wonder why the DKLT is at all necessary: our statistic
in Eq.~\prettyref{eq:statisticWithInverse} could be computed just
inverting the matrix \( \left( \mathbf{R}_{y}\right) _{\parallel } \),
and then applying it to each successive data chunk; what is the advantage
of the expression in Eq.~\prettyref{eq:statisticDKLT}? The answer
is that the computational cost is the same, but the DKLT decomposition
gives us more flexibility. We are not forced to sum over all the elements:
we can decide for instance that some of the basis elements correspond
to large noise components, and can be left out without significantly
affecting the detector performance. The fact that the coefficients
\( \bm \psi _{\parallel }^{k}\cdot \mathbf{y}_{\parallel } \), for
different values of \( k \), are by definition statistically uncorrelated
renders this procedure sound and does not complicate the statistical
analysis .

We will elaborate more on the practical implementation in \prettyref{sec:implementation}:
we now turn to consider two special cases which help building a better
understanding.

\subsubsection{A special case: \protect\( N=N_{\parallel }\protect \)}

This is unrealistic: in this case there is no orthogonal space, \( \left( \mathbf{R}^{-1}_{n}\right) _{\parallel }=\mathbf{R}_{n}^{-1} \)
and the likelihood ratio becomes\begin{equation}
\Lambda (\mathbf{x})=\exp \left[ \frac{1}{2}\mathbf{x}\cdot \mathbf{R}_{n}^{-1}\cdot \mathbf{x}\right] \, ;
\end{equation}
if \( \left\{ \bm \phi ^{k},\, k=1,\, N\right\}  \) is the appropriate
DKL basis for the noise correlation matrix \( \mathbf{R}_{n} \),
that is\begin{equation}
\mathbf{R}_{n}=\sum ^{N}_{k=1}\sigma _{k}\bm \phi ^{k}\otimes \bm \phi ^{k}\, ;
\end{equation}
one would therefore use as a statistic for burst searches\begin{equation}
L=\sum _{k}\frac{1}{\sigma _{k}}\left( \bm \phi ^{k}\cdot \mathbf{x}\right) ^{2}\, .
\end{equation}
This expression corresponds closely to the excess energy statistic
defined in\cite[Eq. (1.4)]{anderson:brady:creighton:flanagan:01},
with two differences: (1) by using the DKL expansion it takes into
account the finite size of the sample; (2) it is shown to be appropriate,
given our assumptions on the signal distribution, only in this very
special case when no information is available \emph{before} and \emph{after}
the data segment we are searching for a burst. The correspondence
makes more evident why our result differs from theirs: having they
chosen a uniform prior with the noise metric, they have effectively
decoupled the \( {\cal V}_{\perp } \), \( {\cal V}_{\parallel } \)
subspaces, which is equivalent in our context to neglecting the presence
of correlated noise.

\subsubsection{\label{sec:approximateStatistic}Approximate expression: \protect\( N_{\parallel }\protect \)
large}

Let us go back to the case in which we search for a burst of length
\( N_{\parallel } \) in a much longer data train of length \( N \):
as we said, sufficiently long to resolve the narrow spectral features
which give rise to long correlation times. The high sampling rate
\( O(20\, \mathrm{kHz}) \) needed to exploit the wide spectral range
available in interferometer data may lead to consider \( N_{\parallel } \)
of the order of several hundred samples, even for signals of a few
tens of ms.

If this results in a excessive computational cost in the application
of the KL transform, we can exploit its convergence to the Fourier
Transform, in the limit of \( N_{\parallel } \) large: namely, we
can write down approximately\begin{equation}
\left[ \left( \mathbf{R}_{y}\right) _{\parallel }\right] ^{-1}\simeq f_{s}^{-1}\sum ^{N_{\parallel }-2}_{k=1}\frac{2}{S_{y}[k]}\mathbf{w}_{k}\otimes \mathbf{w}^{H}_{k}
\end{equation}
where \( \mathbf{w}_{k} \) are the Fourier basis vectors (see \prettyref{sec:conventions})
in \( {\cal V_{\parallel }} \), approximating the DKL transform for
\( \left( \mathbf{R}_{y}\right) _{\parallel } \) with eigenvalues
\( \lambda _{k}=\frac{1}{2}S_{y}\left[ k\right] \, f_{s} \).

It follows the expression for the approximate statistic\begin{eqnarray}
L(\mathbf{x}) & \simeq  & f^{-1}_{s}\sum ^{N_{\parallel }-2}_{k=1}\frac{2}{S_{y}\left[ k\right] }\left| \mathbf{w}_{k}\cdot \mathbf{y}_{\parallel }\right| ^{2}\nonumber \\
 & = & \frac{f_{s}}{N_{\parallel }}\sum ^{N_{\parallel }/2-1}_{k=1}\frac{1}{S_{y}\left[ k\right] }\left| \tilde{y}_{\parallel }\left[ k\right] \right| ^{2};\label{eq:statisticApproximate} 
\end{eqnarray}
\( L\left( \mathbf{x}\right)  \) is the sum of the squares of the
Fourier coefficients \( \tilde{y}_{\parallel }\left[ k\right] \equiv f^{-1}_{s}\sqrt{N_{\parallel }}\, \mathbf{w}_{k}\cdot \mathbf{y}_{\parallel } \)
of the time series \( \mathbf{y}_{\parallel }\equiv \left( \mathbf{R}_{n}^{-1}\cdot \mathbf{x}\right) _{\parallel } \),
weighted with the corresponding spectral noise density. This expression
is similar to the excess energy statistic defined in\cite{flanagan:hughes:98,anderson:brady:creighton:flanagan:01}:
the difference, apart the \( N_{\parallel } \) large approximation,
is that we found it necessary to start from data filtered for the
occurrence of \( \delta  \).

\subsection{\label{sec:discussion}In which sense the statistic is {}``optimal''}

Before moving to the practical implementation of the algorithm, it
is very important to fully understand the consequences of the assumption
we have made about the \emph{a priori} distribution of the signals:
to this end, let us have a second look at the likelyhood ratio\noun{\begin{equation}
\Lambda (\mathbf{x})=\int e^{-\frac{1}{2}\mathbf{s}\cdot \mathbf{R}_{n}^{-1}\cdot \mathbf{s}+\mathbf{s}\cdot \mathbf{R}_{n}^{-1}\cdot \mathbf{x}}P(\mathbf{s})d\mathbf{s}.
\end{equation}
}Our choice has been that \( P\left( \mathbf{s}\right) =1 \), which
is a way to avoid introducing any scale in the problem, which might
bias the analysis. There is however a drawback: larger values of \( \mathbf{s}\cdot \mathbf{s} \)
are favored, in fact\begin{equation}
d\mathbf{s}=\rho ^{N_{\parallel }-1}d\rho d\Omega _{{N_{\parallel }}}\left( \hat{\mathbf{s}}\right) 
\end{equation}
where \( \rho \equiv \sqrt{\mathbf{s}\cdot \mathbf{s}} \) and \( d\Omega _{N_{\parallel }} \)
is the solid angle element in \( N_{\parallel } \) dimensions. If
we have (say) \emph{a priori} information only on the energy of the
signal, or equivalently on the distribution \( p\left( \rho \right)  \),
then we would like to follow the same approach as in \cite[Sec. III]{anderson:brady:creighton:flanagan:01}
and write\begin{equation}
\Lambda \left( \mathbf{x}\right) =\int p\left( \rho \right) \Lambda \left( \mathbf{x}|\rho \right) d\rho 
\end{equation}
where\begin{equation}
\Lambda \left( \mathbf{x}|\rho \right) \equiv \int \delta \left( \rho -\sqrt{\mathbf{s}\cdot \mathbf{s}}\right) e^{-\frac{1}{2}\mathbf{s}\cdot \left( \mathbf{R}_{n}^{-1}\right) _{\parallel }\cdot \mathbf{s}+\mathbf{s}\cdot \mathbf{y}_{\parallel }}d\mathbf{s}
\end{equation}
and \( \mathbf{y}=\mathbf{R}^{-1}_{n}\cdot \mathbf{x} \) belongs
to the parallel space \( {\cal V}_{\parallel } \). Changing basis
with a DKL transform \( \mathbf{s}\rightarrow \bm \varsigma  \),
we obtain\begin{eqnarray}
\Lambda \left( \mathbf{x}|\rho \right)  & = & \int \delta \left( \rho -\sqrt{\bm \varsigma \cdot \bm \varsigma }\right) e^{-\frac{1}{2}\sum _{k}\sigma _{k}\varsigma ^{2}_{k}+\bm \varsigma \cdot \mathbf{c}}d\bm \varsigma \nonumber \\
 & = & \int e^{-\frac{\rho ^{2}}{2}\sum _{k}\sigma ^{2}_{k}\hat{\varsigma }^{2}_{k}+\rho \sum _{k}\hat{\varsigma }_{k}c_{k}}d\Omega _{N_{\parallel }}\left( \hat{\varsigma }\right) \label{eq:lambdaXrho} 
\end{eqnarray}
where \( c_{k}\equiv \bm \psi _{k}\cdot \mathbf{y}_{\parallel } \)
and \( \sigma _{k} \) are the eigenvalues of \( \left( \mathbf{R}^{-1}_{n}\right) _{\parallel } \).
If we had \( \sigma _{k} \) independent on \( k \), in other words
if all the directions in the \( \bm \varsigma  \) space were equivalent,
we could as in \cite[Sec. III]{anderson:brady:creighton:flanagan:01}
compute the integral in closed form, by aligning one of the axes with
the direction of the vector \( \mathbf{c} \). This was possible to
Anderson \emph{et al.} because they had chosen a signal prior function
of \( \mathbf{s}\cdot \left( \mathbf{R}_{n}\right) ^{-1}_{\parallel }\cdot \mathbf{s} \).

In the case considered by us the expression in Eq.\prettyref{eq:lambdaXrho}
is not in general a function solely of the statistic \( L=\sum _{k}\sigma ^{-1}_{k}c^{2}_{k} \)
and of \( \rho  \), because the noise introduces preferential directions
in the space of possible signals.

This discussion shows that the statistic \( L \) we have proposed
is strictly speaking \emph{optimal} only for a signal prior \( P\left( \mathbf{s}\right)  \)
constant%
\footnote{The statistic \( L \) would of course be optimal also for any prior
which is itself a function of the statistic \( L \).
}: we can make the \emph{ansatz} that for a more general prior, depending
on a scalar function of \( \mathbf{s} \), the optimal statistic \( L \)
might still be of the form\begin{equation}
L=\sum _{k,l}\mathsf{L}_{kl}\frac{c_{k}c_{l}}{\sqrt{\sigma _{k}\sigma _{l}}}
\end{equation}
where the matrix \( \mathsf{L} \) should be determined maximizing
the probability of detection while keeping the false alarm rate fixed.
We were however unable to prove that this is actually the case, at
least under certain restrictions on the form of \( P\left( \mathbf{s}\right)  \).

Another implication of this discussion is that with our statistic
it is difficult to set Bayesian thresholds, as derived by choosing
a particular form for \( p\left( \rho \right)  \), and making assumptions
on its parameters. This is however a somewhat less crucial issue because
thresholds may be set following a frequentist approach, that is by
limiting the false alarm rate, as shall be discussed in \prettyref{sec:statisticalAnalysis}.

\subsection{\label{sec:implementation}Description of the algorithm}

The expressions in Eqs.~(\ref{eq:exactStatistic},\ref{eq:statisticDKLT},\ref{eq:statisticApproximate})
define the algorithm for estimating the (log)-likelihood, at different
approximation levels: we find useful to detail the procedure.

We assume that the starting point is a continuous stream of data,
at sampling frequency \( f_{s} \), whose Gaussian noise component
is assumed to be stationary. The purpose of the algorithm is to search
the stream for bursts of length \( N_{\parallel } \) and unknown
shape; the choice of \( N_{\parallel } \) is arbitrary if there is
no physical hint.

The steps of the algorithm are the following:

\begin{enumerate}
\item partition the data stream into data vectors \( \mathbf{x} \) of length
\( N\gg N_{\parallel } \); each vector should overlap the following
by a certain amount \( 2M \) which needs tuning (it is related to
boundary effects on the correlation matrix, which depend on the specific
noise considered). The length \( N \) must be sufficient to resolve
the narrow spectral features in the data, or in other words to ensure
that the correlation function \( R_{n}\left( l/f_{s}\right) =E\left[ x[i]\, x[i+l]\right]  \)
is sampled over a sufficiently long interval, allowing it to decrease
to zero with sufficient accuracy.
\item Estimate the correlation matrix, or equivalently the \emph{sample}
spectrum \( S_{n}[k] \), over the \( \mathbf{x} \) segments: that
is, with frequency resolution \( df=\frac{f_{s}}{N} \). For instance
using a Welch's overlap and save procedure to combine estimates from
different vectors of length \( N \), or averaging estimates obtained
with a multi-taper spectral estimator\cite{percival:walden:93}. The
cost of this step, per vector, is \( O(N\ln N) \).
\item For each data vector \( \mathbf{x} \), estimate a new vector \( \mathbf{y}\equiv \mathbf{R}_{n}^{-1}\cdot \mathbf{x} \);
this is equivalent, as we have seen in \prettyref{sec:burstUnit},
to filter for \( \delta  \) functions. It can be done at least in
two ways, not necessarily equivalent from the numerical point of view:

\begin{enumerate}
\item in the frequency domain, Fourier transforming the data \( \mathbf{x}\rightarrow \tilde{\mathbf{x}} \)
and defining\begin{equation}
\label{eq:deltaFilterDiscrete}
y[l]=\frac{1}{N}\sum ^{N-1}_{k=0}\frac{2}{S_{n}\left[ k\right] }e^{-i\, 2\pi \, k\, l/N}\tilde{x}[k]\, ;
\end{equation}
 the computational cost is again \( O(N\ln N) \).
\item In the time domain, first performing a spectral factorization \( \mathbf{R}_{n}^{-1}=\mathbf{W}_{n}^{\mathbf{t}}\cdot \mathbf{W}_{n} \)
which defines two matrix operators \( \mathbf{W}_{n} \) and \( \mathbf{W}_{n}^{\mathbf{t}} \),
the first causal and the other anti-causal. Both can be implemented
as digital filters: the anti-causal one after reversing the input
data. Several methods exist to perform such a factorization and to
apply the resulting filters to the data \cite{papoulis:91,therrien:92}:
see \cite{cuoco:01,cuoco:02} for an application to the analysis of
data from the Caltech 40m detector \cite{ligo40mData:96}. If the
noise is stationary across several data vectors, the computational
cost of the spectral factorization itself becomes negligible: otherwise,
it can be shown \cite{therrien:92} that the cost of estimating filters
with \( P \) coefficients, using for instance the Levinson-Durbin
recursion, grows as \( P^{2} \), where the right order depends on
the spectral characteristics of the data (see \cite{cuoco:02}, where
it has been shown that FIR filters a few hundred taps are sufficient
for correctly whitening real interferometer data). Once the filters
are estimated, they can be applied with a cost \( O(N\, P) \). One
possible advantage of these time-domain methods, which are generally
slower than the FFT (but can be implemented very efficiently on DSP
systems) is that they can follow a slow noise non-stationarity: this
issue however requires more study, because the very definition of
the {}``burst'' statistic needs a revision, when the noise is not
stationary.
\end{enumerate}
\item Drop the first and last \( M \) points in the data vector \( \mathbf{y} \),
in order to reduce the boundary effects, and partition it in segments
\( \mathbf{y}_{\parallel } \) of length \( N_{\parallel } \). The
partitioning will require to overlap the \( \mathbf{y}_{\parallel } \)
segments: in any case one will have to take into account the correlation
of the resulting statistics.
\item Use the \( N-2M \) data in \( \mathbf{y} \) to estimate the correlation
matrix \( \mathbf{R}_{y} \) over lags of at most \( N_{\parallel } \):
this operation can be done using DFT methods with a cost at most \( O\left( N\times \ln N\right)  \),
or smaller if we can exploit the fact that only \( \left( \mathbf{R}_{y}\right) _{\parallel } \)
is needed.
\item Decompose of the matrix \( \left( \mathbf{R}_{y}\right) _{\parallel } \),
using either the Karhunen-Lo\`eve or the Fourier transform.

\begin{enumerate}
\item \label{step:DKLTdecomposition}In the case of the DKLT, one needs
the eigenvalues \( \sigma _{k} \) and eigenvectors \( \bm \psi _{\parallel }^{k} \)
of the matrix\begin{equation}
\left( \mathbf{R}_{y}\right) _{\parallel }=\sum ^{N_{\parallel }}_{k=1}\sigma _{k}\, \bm \psi _{\parallel }^{k}\otimes \bm \psi _{\parallel }^{k}
\end{equation}
 this decomposition has to be done at most once for each \( \mathbf{y} \)
vector, and possibly even more rarely, depending on the noise stationarity.
Generally the cost of this decomposition is \( O(N^{3}_{\parallel }) \),
unless one is able to exploit the Toeplitz structure of the matrix
\( \mathbf{R}_{y} \). One possibly selects then only a subset of
the eigenvectors, chosen for instance by setting a threshold on the
value of the eigenvalues \( \sigma _{k} \). This subset consists
of \( N_{KL}\leq N_{\parallel } \) elements, to be used in the actual
evaluation of the statistic. 
\item With the Fourier transform method (provided it is a reasonable approximation)
we need instead only the spectrum of the time series \( \mathbf{y} \),
at a reduced frequency resolution \( df_{\parallel }\equiv f_{s}/N_{\parallel } \):
we call this spectrum \( {S_{y}[k],\, k\in [0,\, N_{\parallel }-1]} \):
it can also be computed by averaging the bins of the \( \left( S_{n}\right) ^{-1} \),
if the noise is stationary across instances of the vector \( \mathbf{x} \).
\end{enumerate}
\item For each segment \( \mathbf{y}_{\parallel } \) assemble the statistic
for the log likelihood, following a recipe dependent on the chosen
method:

\begin{enumerate}
\item using the DKLT one computes\begin{equation}
\label{eq:defStatisticL}
L=\sum ^{N_{KL}}_{k=1}\frac{1}{\sigma _{k}}\left[ \bm \psi ^{k}\cdot \mathbf{y}_{\parallel }\right] ^{2},
\end{equation}
 with a cost \( O(N_{KL}N_{\parallel })\leq O(N^{2}_{\parallel }) \)
in floating-point operations: this recipe gives the exact result. 
\item Using the Fourier transform one instead first computes the Fourier
coefficients \( \tilde{\mathbf{y}}_{\parallel }\left[ k\right]  \),
a operation costing \( O(N_{\parallel }\times \ln N_{\parallel }) \):
then one combines the outputs in the statistic\begin{equation}
\label{eq:defStatisticLapproximate}
L=\frac{f_{s}}{N_{\parallel }}\sum ^{N_{\parallel }/2-1}_{k=1}\frac{1}{S_{y}[k]}\left| \tilde{\mathbf{y}}_{\parallel }\left[ k\right] \right| ^{2}\, ;
\end{equation}
 this expression is approximate, but its accuracy increases with \( N_{\parallel } \),
and can be legitimate in some cases. 
\end{enumerate}
\item Perform the statistical analysis of the results: we will discuss about
this step in \prettyref{sec:statisticalAnalysis}.
\end{enumerate}
A few comments are in order:

\begin{itemize}
\item even though the procedure is designed to cope with long correlations
in the data, we have implicitly assumed that no \emph{deterministic}
lines are present: in fact, the Wold Decomposition theorem\cite[sec. 7.6.2]{therrien:92}
states that a general random process can be decomposed in the sum\begin{equation}
x(t)=x_{r}(t)+x_{p}(t)
\end{equation}
 of a \emph{regular} process%
\footnote{A regular random process \( x[l] \) satisfies the Paley-Wiener condition\cite[sec. 5.5.2]{therrien:92}
$\int_{-\pi}^\pi\left|\ln S_x[\omega]\right|~d\omega < \infty$ and can be written as white noise filtered through a causal process: this is usually the case for physical noises.
} \( x_{r}(t) \) and a \emph{predictable} process%
\footnote{A process is said predictable if the output at time \( t+dt \) can
be written in terms of past values \emph{without} error: such a process
clearly cannot be written as white noise filtered through a causal
filter\cite[sec. 7.6.1]{therrien:92}.
} \( x_{p}(t) \). The latter could correspond to a harmonic of the
power line: it would contribute to the spectrum a term\begin{equation}
\sigma ^{2}_{p}\delta \left( \nu -\nu _{p}\right) 
\end{equation}
where \( \nu _{p} \) is the frequency of the line and \( \sigma _{p} \)
its contribution to the RMS noise. In the sample spectrum \( S_{n}\left[ k\right]  \)
this feature would translate approximately into a term\begin{equation}
\frac{N}{f_{s}}\, \delta _{k,k_{p}}=T\, \delta _{k,k_{p}},
\end{equation}
a trend in \( T \) which is the symptom of a infinite correlation
length. Such spectral features cannot be properly handled by spectral
factorization methods, because they are deterministic components and
not stochastic. It is advisable to subtract altogether such predictable
components, because this operation can certainly improve the signal
to-noise ratio : several examples are available in literature \cite{sintes:schutz:98,flaminio:00}.
Notice \emph{in passim} that violin mode lines excited by thermal
noise belong to the class of regular processes, although they can
also be modeled and partially subtracted \cite{finn:mukherjee:01}.
\item If the noise is not stationary, we can still perform an adaptive whitening,
and refresh the estimate of the DKL basis \textbf{\( \{\bm \psi ^{k}\} \)}
each time it is found necessary. However as we have anticipated, further
study is necessary; for instance, the very definition of \( \mathbf{R}_{y} \)
as an average becomes uncertain when one cannot trade ensemble averaging
for time averaging. In case of a slow non-stationarity one may also
think of using an adaptive estimator of the \( \delta  \) filter
in Eq.~\prettyref{eq:deltaFilterDiscrete}; this is a good topic
for future research work.
\item Whether it is necessary to adopt the exact statistic in Eq.~\prettyref{eq:defStatisticL},
or the approximate one in Eq.~\prettyref{eq:defStatisticLapproximate},
can only be judged comparing the receiver performance in a definite
situation.
\end{itemize}
We turn out to the analysis of the distribution of our statistic for
burst detection.

\subsection{\label{sec:statisticalAnalysis}Statistical analysis}

Given the statistic \( L \) we need the distributions \( d_{0}\left( L\right)  \)
and \( d_{1}\left( L|\mathbf{s}\right)  \) under the respective hypotheses
\( H_{0} \) (no signal) and \( H_{1} \) (a signal \( \mathbf{s} \)
of unspecified form), in order to set up a detection strategy, for
instance based on the Neyman-Pearson criterion \cite{helstrom:60}.

The distribution \( d_{0}\left( L\right)  \) is immediately recognized
to be a \( \chi ^{2}\left( N_{\parallel }\right)  \); we find illustrative
to prove this directly, from the definition

\begin{eqnarray}
d_{0}(L) & \propto  & \int e^{-\frac{1}{2}\mathbf{n}\cdot \mathbf{R}_{n}^{-1}\cdot \mathbf{n}}\\
 & \times  & \delta \left( L-\left( \mathbf{R}_{n}^{-1}\mathbf{n}\right) _{\parallel }\left( \left( \mathbf{R}^{-1}_{n}\right) _{\parallel }\right) ^{-1}\left( \mathbf{R}_{n}^{-1}\mathbf{n}\right) _{\parallel }\right) d\mathbf{n}\nonumber \\
 & \propto  & \int e^{-\frac{1}{2}\mathbf{y}\left( \mathbf{R}_{y}\right) ^{-1}\mathbf{y}}\delta \left( L-\mathbf{y}_{\parallel }\left( \left( \mathbf{R}_{y}\right) _{\parallel }\right) ^{-1}\mathbf{y}_{\parallel }\right) d\mathbf{y}_{\parallel }d\mathbf{y}_{\perp }\nonumber 
\end{eqnarray}
and observing that\begin{equation}
\label{eq:formRyInverse}
\mathbf{R}_{y}^{-1}=\left( \begin{array}{cc}
\left( \mathbf{R}_{y}\right) _{\parallel } & \mathbf{B}\\
\mathbf{B}^{t} & \left( \mathbf{R}_{y}\right) _{\perp }
\end{array}\right) ^{-1}=\left( \begin{array}{cc}
\mathbf{A} & \mathbf{D}\\
\mathbf{D}^{t} & \mathbf{C}
\end{array}\right) 
\end{equation}
where\begin{subequations}\begin{eqnarray}
\mathbf{A} & = & \left( \left( \mathbf{R}_{y}\right) _{\parallel }\right) ^{-1}+\nonumber \\
 &  & +\left[ \left( \left( \mathbf{R}_{y}\right) _{\parallel }\right) ^{-1}\mathbf{BCB}^{t}\left( \left( \mathbf{R}_{y}\right) _{\parallel }\right) ^{-1}\right] \label{eq:formAmatrix} \\
\mathbf{D} & = & -\left( \left( \mathbf{R}_{y}\right) _{\parallel }\right) ^{-1}\mathbf{BC}
\end{eqnarray}
\end{subequations}and the explicit form of \( \mathbf{C} \) is not
relevant for us. Integrating over \( \mathbf{y}_{\perp } \) one obtains
a factor\begin{eqnarray}
\int d\mathbf{y}_{\perp }e^{-\frac{1}{2}\mathbf{y}_{\perp }\mathbf{C}\, \mathbf{y}_{\perp }+\mathbf{y}_{\perp }\mathbf{C}\, \mathbf{B}^{t}\left( \left( \mathbf{R}_{y}\right) _{\parallel }\right) ^{-1}\mathbf{y}_{\parallel }} & \propto  & \nonumber \\
e^{\frac{1}{2}\left[ \left( \left( \mathbf{R}_{y}\right) _{\parallel }\right) ^{-1}\mathbf{B}\, \mathbf{C}\, \mathbf{B}^{t}\, \left( \left( \mathbf{R}_{y}\right) _{\parallel }\right) ^{-1}\right] } &  & ;
\end{eqnarray}
 comparing with Eqs.~(\ref{eq:formRyInverse},\ref{eq:formAmatrix}),
the term in square parentheses is recognized to be \( \left( \mathbf{R}_{y}^{-1}\right) _{\parallel }-\left( \left( \mathbf{R}_{y}\right) _{\parallel }\right) ^{-1} \),
hence\begin{eqnarray}
d_{0}\left( L\right)  & \propto  & \int e^{-\frac{1}{2}\mathbf{y}_{\parallel }\left( \left( \mathbf{R}_{y}\right) _{\parallel }\right) ^{-1}\mathbf{y}_{\parallel }}\nonumber \\
 &  & \times \delta \left( L-\mathbf{y}_{\parallel }\left( \left( \mathbf{R}_{y}\right) _{\parallel }\right) ^{-1}\mathbf{y}_{\parallel }\right) d\mathbf{y}_{\parallel }\nonumber \\
 & = & \frac{L^{N_{\parallel }/2-1}}{2^{N_{\parallel }/2}\Gamma (N_{\parallel }/2)}\, e^{-L/2}\label{eq:chiSquareN} 
\end{eqnarray}
which is as expected the distribution of a \( \chi ^{2}\left( N_{\parallel }\right)  \)
\cite[Sec. 4-3]{papoulis:91}: we underline the role of the cross
correlation effects, in order to derive the correct result.

\begin{widetext}

When on the contrary a signal is present, the distribution can be
written as 

\begin{eqnarray}
d_{1}\left( L|\mathbf{s}\right)  & = & {\cal N}\int e^{-\frac{1}{2}\mathbf{n}\cdot \mathbf{R}_{n}^{-1}\cdot \mathbf{n}}\delta \left( L-\left( \mathbf{R}_{n}^{-1}\cdot \left( \mathbf{n}+\mathbf{s}\right) \right) _{\parallel }\cdot \left( \left( \mathbf{R}^{-1}_{n}\right) _{\parallel }\right) ^{-1}\cdot \left( \mathbf{R}_{n}^{-1}\cdot \left( \mathbf{n}+\mathbf{s}\right) \right) _{\parallel }\right) \, d\mathbf{n}\nonumber \\
 & = & {\cal N}\int e^{-\frac{1}{2}\left( \mathbf{y}-\mathbf{R}^{-1}_{n}\cdot \mathbf{s}\right) _{\parallel }\cdot \left( \left( \mathbf{R}_{y}\right) _{\parallel }\right) ^{-1}\cdot \left( \mathbf{y}-\mathbf{R}^{-1}_{n}\cdot \mathbf{s}\right) _{\parallel }}\delta \left( L-\mathbf{y}_{\parallel }\cdot \left( \left( \mathbf{R}_{y}\right) _{\parallel }\right) ^{-1}\cdot \mathbf{y}_{\parallel }\right) d\mathbf{y}_{\parallel }
\end{eqnarray}
which immediately results in the non-central \( \chi ^{2}\left( N_{\parallel }\right)  \):
in terms of the signal-to-noise ratio (\textsf{SNR)} one has\begin{equation}
\label{eq:P1complete}
d_{1}(L|\mathsf{SNR})=\frac{L^{N_{\parallel }/2-1}}{2^{N_{\parallel }/2}\Gamma (N_{\parallel }/2)}e^{-\frac{1}{2}\left( L+\sqrt{2N_{\parallel }}\, \mathsf{SNR}\right) }\, _{0}\mathrm{F}_{1}\left( ;\frac{N_{\parallel }}{2};\frac{\mathsf{SNR}\, L\sqrt{2N_{\parallel }}}{4}\right) ,
\end{equation}
where \( _{0}F_{1} \) is a hyper-geometric function\cite[sec. 9.14]{gradshteyn:ryzhik:65}.%
\footnote{It can be expressed in terms of $\Gamma$ and Bessel functions $\mathrm{I}$
as follows: $_0F_1(;x,y)=y^{\frac{1-x}{2}}\Gamma(x)\ \mathrm{I}_{x-1}\left(2\sqrt{y}\right)$
}

\end{widetext}

The \textsf{SNR} explicit form is \begin{equation}
\mathsf{SNR}=\frac{\left( \mathbf{R}^{-1}_{n}\cdot \mathbf{s}\right) _{\parallel }\cdot \left( \left( \mathbf{R}_{y}\right) _{\parallel }\right) ^{-1}\cdot \left( \mathbf{R}^{-1}_{n}\cdot \mathbf{s}\right) _{\parallel }}{\sqrt{2N_{\parallel }}}
\end{equation}
consistently with the general definition \cite[chap. 6]{donough:whalen:95}:
\begin{equation}
\label{eq:SNRdefinition}
\mathsf{SNR}\equiv \frac{\left| E[L|H_{1}]-E[L|H_{0}]\right| }{\sqrt{E\left[ \left( L-E[L|H_{0}]\right) ^{2}|\, H_{0}\right] }}\, ,
\end{equation}
 where \( E\left[ L|H\right]  \) is the expectation value of the
statistic \( L \) under the hypothesis \( H \), and \( H_{1} \),
\( H_{0} \) correspond respectively to the hypotheses of presence
of absence of a signal. Please note that the \textsf{SNR} defined
in Eq.~\prettyref{eq:SNRdefinition} has nothing to do with the \emph{intrinsic}
signal-to-noise ratio which would result from a matched filter procedure,
\( \mathsf{SNR}_{\mathsf{intrinsic}}\propto \sqrt{\int \left| \tilde{s}\left( f\right) \right| ^{2}/S_{n}\left( f\right) \, df} \);
in particular, our \textsf{SNR} is quadratic in the signal amplitude.

Given the distribution \( d_{0}\left( L\right)  \) the false alarm
probability can be readily computed as\begin{equation}
\label{eq:Qf}
Q_{f}(L_{0})\equiv \int ^{\infty }_{L_{0}}d_{0}\left( L\right) \, dL=\frac{\Gamma \left( \frac{N_{\parallel }}{2},\frac{L_{0}}{2}\right) }{\Gamma \left( \frac{N_{\parallel }}{2}\right) }\, ,
\end{equation}
while the detection probability\begin{equation}
Q_{d}\left( L_{0}|\mathsf{SNR}\right) \equiv \int ^{\infty }_{L_{0}}d_{1}\left( L|\mathsf{SNR}\right) \, dL
\end{equation}
cannot be written in closed form. For large \textsf{SNR} it is possible
to approximate \( d_{1}\left( L|\mathsf{SNR}\right)  \) with a Gaussian
distribution having the same first and second order momenta\begin{equation}
\label{eq:P1largeSNR}
d_{1}(L|\mathsf{SNR})\simeq \frac{e^{-\frac{\left( L-(\mathsf{SNR}\sqrt{2N_{\parallel }}+N_{\parallel })\right) ^{2}}{4(N_{\parallel }+2\, \mathsf{SNR}\sqrt{2N_{\parallel }}}}}{\sqrt{4\pi (N_{\parallel }+2\, \mathsf{SNR}\sqrt{2N_{\parallel }})}}
\end{equation}
obtaining\begin{equation}
\label{eq:Qd}
Q_{d}\simeq \frac{1}{2}\left[ 1+\mathrm{erf}\left( \frac{N_{\parallel }+\mathsf{SNR}\sqrt{2N_{\parallel }}-L_{0}}{2\sqrt{N_{\parallel }+2\, \mathsf{SNR}\sqrt{2N_{\parallel }}}}\right) \right] \, .
\end{equation}
 The expressions for \( Q_{f} \) and \textbf{\( Q_{d} \)} are the
building blocks for setting up the detection strategy: they are appropriate
if we are able to compute the exact statistic \( L \). How should
however \( d_{0} \) and \( d_{1} \) be modified, if we have instead
chosen to compute an approximate statistic?

\subsubsection{\label{sec:approximateStatisticDistribution}Approximate statistic
distribution}

Recall that we have discussed in \prettyref{sec:exactStatistic} also
the possibility of defining the statistic \( L \), when computed
using the DKL expansion, using just a subset \( N_{KL}<N_{\parallel } \)
of the basis vectors: assuming to have ordered the \( \bm \psi ^{k} \)
vectors and taking only the first \( N_{KL} \), we define\begin{equation}
L=2\ln \Lambda (\mathbf{n})=\sum ^{N_{KL}}_{k=1}\frac{1}{\sigma _{k}}(\bm \psi ^{k}\cdot \left( \mathbf{R}_{n}^{-1}\cdot \mathbf{n}\right) _{\parallel })^{2}\, ,
\end{equation}
where we dropped the suffix \( \parallel  \) from the basis vectors
\( {\bm \psi ^{k}} \); recall that the basis is independent on the
specific \( N_{\parallel } \) segment in the data train. We know
that in absence of signals the expansion coefficients \( \xi _{k}=\frac{1}{\sqrt{\sigma _{k}}}\bm \psi ^{k}\cdot \left( \mathbf{R}^{-1}\cdot \mathbf{n}\right) _{\parallel } \)
are by construction uncorrelated, with zero mean and unit variance
(see Eq.~\prettyref{eq:avDKLTcoeff}); they are also Gaussian variables,
because they are linear combinations of Gaussian variables. Hence
\( L \) is distributed as a \( \chi ^{2}\left( N_{KL}\right)  \),
and analogously the formula in Eq.~\prettyref{eq:P1complete} for
\( d_{1}\left( L|\mathsf{SNR}\right)  \) holds, where \( N_{\parallel }\rightarrow N_{KL} \)
and\begin{equation}
\mathsf{SNR}\equiv \frac{1}{\sqrt{2\, N_{KL}}}\sum ^{N_{KL}}_{k=1}\frac{1}{\sigma _{k}}\left[ \bm \psi ^{k}\cdot \left( \mathbf{R}_{n}\cdot \mathbf{s}\right) _{\parallel }\right] ^{2}\, ;
\end{equation}
again, these results are exact, despite the fact that we are using
only a subset of the DKL vectors. This might be useful in order to
implement \( \chi ^{2} \) tests for non-Gaussianity, similarly to
what has been done in the analysis of Caltech 40m data while searching
for coalescing binaries signals \cite{allen-et-al:99}.

The other possible approximation is the use of the Fourier transform
instead of the DKL transform, as discussed in \prettyref{sec:approximateStatistic}.
In the limit in which this approximation is legitimate, namely, when
\( N_{\parallel } \) is large and the \( \bm \psi ^{k} \) vectors
converge to the vectors \( \mathbf{w}^{k} \) of the Fourier basis,
the coefficients of the Fourier expansion behave from the statistical
point of view as those of the DKL expansion, and the same results
apply.

\section{\label{sec:multipleDetector}Multiple detector case}

\subsection{The signal at each detector}

The mathematical tools needed to compactly describe the response of
interferometric detectors to a coherent GW signal have been laid out
in several papers \cite{gelfand:63,thorne:80,dhurandhar:tinto:88},
and have been recently reviewed and applied to the problem of network
detection of coalescing binary signals \cite{pai:dhurandhar:bose:01}.
We collect the definitions and formulas useful to us in \prettyref{sec:geometry}
and we refer particularly to \cite{pai:dhurandhar:bose:01}, whose
nomenclature we follow closely, for a complete treatment.

We describe the incoming gravitational wave by means of a \emph{wave
frame,} having the Z \textbf{}axis aligned with the direction of propagation;
the signal is parameterized by two polarizations \( h_{+,\times } \),
whose definition depends on the orientation of the \textbf{X}, \textbf{Y}
axes; as in \prettyref{sec:burstIntegration}, we will regard \( h_{+,\times } \)
as independent nuisance parameters.

Another important frame is the \emph{network frame}, defined as centered
on the Earth and having the \textbf{Z} axis oriented along the North
pole and the \textbf{X} axis along the Greenwich meridian: rotations
between the network frame and the wave frame are accomplished by the
Euler angles \( \phi ,\theta ,\psi  \); we can set \( \psi =0 \)
from now on, because it specifies a rotation of the wave frame around
the \textbf{Z} axis, which is inessential. In turn we can define reference
frames centered on the detectors (see \prettyref{sec:geometry}),
and call \( \alpha _{\mathsf{L}},\beta _{\mathsf{L}},\gamma _{\mathsf{L}} \)
the Euler angles needed to rotate between the network frame and the
\emph{detector frame} relative to the \textsf{L-}th \textsf{}detector:
see \prettyref{sec:networkCharacteristics} for the values of these
angles for the interferometers under construction.

We write therefore the signal at the \textsf{L}-th detector as\begin{equation}
\label{eq:signalAtDetector}
s_{\mathsf{L}}\left( t\right) =h_{+}\left( t-\tau _{\mathsf{L}}\right) \, F^{+}_{\mathsf{L}}+h_{\times }\left( t-\tau _{\mathsf{L}}\right) \, F^{\times }_{\mathsf{L}}
\end{equation}
where \( \tau _{\mathsf{L}} \) is the delay of the signal with respect
to what would be received by a detector at the center of the Earth:
it depends on the direction of the source. The two antenna patterns
\( F^{+}_{\mathsf{L}},F^{\times }_{\mathsf{L}} \) are given in \prettyref{sec:geometry}
as functions of the \( \phi ,\theta  \) and \( \alpha _{\mathsf{L}},\beta _{\mathsf{L}},\gamma _{\mathsf{L}} \)
angles; we concentrate on burst signals and omit the dependence on
time of the source location. In other words, the angles \( \phi ,\theta  \)
are a function of time and of declination and right ascension, while
the network frame rotates with the Earth.

Please note that \( \phi ,\theta  \) should not be confused with
the elevation and azimuth angles \( \theta _{s},\phi _{s} \) which
locate the source in polar coordinates with respect to the network
frame: the two sets of angles are related by\begin{equation}
\phi _{s}=\frac{\pi }{2}+\phi ,\, \, \theta _{s}=\pi -\theta \, ;
\end{equation}
in the following we will \emph{always} use, unless explicity stated,
the Euler angles \( \phi ,\theta  \).

\subsection{\label{sec:networkUncorrelatedNoise}Network likelihood with uncorrelated
noise}

We consider first a simpler case, assuming that the Gaussian noise
of the individual detectors is uncorrelated. This would be definitely
true if the only noise sources were the fundamental ones, namely those
dictating the baseline sensitivity (with the possible exception of
the seismic noise for detectors at the same site); technical noises,
like power line interferences \cite{lazzarini:vicere:01}, or external
noises, like correlated magnetic field fluctuations \cite{christensen:90},
may spoil this assumption.

Given this simplification, the likelihood ratio is just the product
of the ratios for the \( M \) individual detectors\begin{equation}
\Lambda \left( \bm x|\, \mathsf{h}\right) \equiv \prod ^{M}_{\mathsf{L}=1}\Lambda _{\mathsf{L}}\left( \mathbf{x}_{\mathsf{L}}|\, \mathbf{s}\right) 
\end{equation}
where we borrow from \cite{finn:01} a bold-italic notation for the
direct sum \( \bm x\equiv \mathbf{x}_{1}\oplus \mathbf{x}_{2}\oplus \cdots \oplus \mathbf{x}_{M} \)
of the data vector from the individual detectors. \( \Lambda  \)
is conditioned by the presence of the signal \( \mathsf{h} \), described
in each detector by Eq.~\prettyref{eq:signalAtDetector} in terms
of the same two polarizations (in the wave frame) \( h_{+,\times } \):
we have\begin{equation}
\Lambda _{\mathsf{L}}\left( \mathbf{x}_{\mathsf{L}}|\, \mathbf{s}\right) \equiv e^{-\frac{1}{2}(\mathbf{s}_{\mathsf{L}})_{i}(\mathbf{R}_{\mathsf{LL}}^{-1})_{ij}(\mathbf{s}_{\mathsf{L}})_{j}+(\mathbf{s}_{\mathsf{L}})_{i}(\mathbf{R}^{-1}_{\mathsf{LL}}\cdot \mathbf{x}_{\mathsf{L}})_{i+d_{\mathsf{L}}}}
\end{equation}
where we have exploited the time invariance of the correlation matrices,
and introduced a shift \( d_{\mathsf{L}} \) in the index of the data
\( \mathbf{x}_{\mathsf{L}} \), changing the reference time at detector
\( \mathsf{L} \) in order to compensate for the delay \( \tau _{\mathsf{L}}\left( \theta ,\phi \right)  \);
hence the burst signal appears at the same time in the individual
data vectors%
\footnote{We are also assuming that all the detectors share data re-sampled
at the same frequency.
}. The matrix \( \mathbf{R}_{\mathsf{LL}} \) represents the noise
autocorrelation for detector \( \mathsf{L} \) and the double index
is irrelevant for the time being, but will be useful when dealing
with cross-detector correlated noise. We can express \( \mathbf{s}_{\mathsf{L}} \)
in terms of the two polarizations, treated as independent vector variables:
it is convenient to write\begin{equation}
\label{eq:signalVector}
s_{\mathsf{L}}\left[ i\right] =\mathsf{h}^{t}\left[ i\right] \cdot \mathsf{F}_{\mathsf{L}}=\left( h_{+}\left[ i\right] ,\, h_{\times }\left[ i\right] \right) \cdot \left( \begin{array}{c}
F^{+}_{\mathsf{L}}\\
F^{\times }_{\mathsf{L}}
\end{array}\right) \, ;
\end{equation}
hence \( \mathsf{h} \) should be regarded as a two column matrix,
and \( \mathbf{s}_{\mathsf{L}} \) is a vector resulting by contracting
one of the indices with those in the vector \( \mathsf{F}_{\mathsf{L}} \).

The likelihood results to be\begin{equation}
\Lambda \left( \bm x|\, \mathsf{h}\right) =e^{-\frac{1}{2}\mathsf{h}^{t}\cdot \left( \sum _{\mathsf{L}}\mathsf{F}_{\mathsf{L}}\otimes \mathbf{R}_{\mathsf{LL}}^{-1}\otimes \mathsf{F}^{t}_{\mathsf{L}}\right) \cdot \mathsf{h}+\mathsf{h}^{t}\cdot \left[ \sum _{\mathsf{L}}\mathsf{F}_{\mathsf{L}}\otimes \mathbf{y}_{\mathsf{L}}\right] }
\end{equation}
where we have introduced the \( \delta  \)-filtered data \begin{equation}
y_{\mathsf{L}}\left[ i\right] =\left( \mathbf{R}^{-1}_{\mathsf{LL}}\cdot \mathbf{x}_{\mathsf{L}}\right) _{i+d_{\mathsf{L}}}
\end{equation}
 including the \( d_{\mathsf{L}} \) time shift.

\begin{widetext}

The Gaussian integration over the \textbf{\( h_{+},\, h_{\times } \)}
vectors can be performed, and we obtain the network log-likelihood\begin{equation}
\label{eq:networkLogLikelyhood}
2\ln \Lambda \left( \bm x|\theta ,\phi \right) =\left[ \sum _{\mathsf{L}}\mathsf{F}_{\mathsf{L}}\otimes \mathbf{y}_{\mathsf{L}}\right] ^{t}_{\parallel }\cdot \left[ \sum _{\mathsf{L}}\mathsf{F}_{\mathsf{L}}\otimes \left( \mathbf{R}_{\mathsf{LL}}^{-1}\right) _{\parallel }\otimes \mathsf{F}^{t}_{\mathsf{L}}\right] ^{-1}\cdot \left[ \sum _{\mathsf{L}}\mathsf{F}_{\mathsf{L}}\otimes \mathbf{y}_{\mathsf{L}}\right] _{\parallel }\, ;
\end{equation}
as expected, the fact that the signal is coherent across detectors
makes it impossible to factor out the integrated likelihood in a product
of terms. The expression obtained is similar to the one proposed by
Anderson \emph{et al}.\cite[Eq. (5.29)]{anderson:brady:creighton:flanagan:01}
apart the fact that, as in the case of a single detector, they have
chosen the signal prior flat in the metric induced by the matrix \( \left[ \sum _{\mathsf{L}}\mathsf{F}_{\mathsf{K}}\otimes \left( \mathbf{R}_{\mathsf{LL}}^{-1}\right) _{\parallel }\otimes \mathsf{F}^{t}_{\mathsf{LL}}\right] ^{-1} \).

\end{widetext}

The formal notation adopted in Eq.~\prettyref{eq:networkLogLikelyhood}
deserves some clarifications:

\begin{description}
\item [\( \mathsf{y}\equiv \begin{array}{cc}
\left( \begin{array}{c}
\mathsf{y}_{+}\\
\mathsf{y}_{\times }
\end{array}\right) =
\end{array}\sum _{\mathsf{L}}\mathsf{F}_{\mathsf{L}}\otimes \mathbf{y}_{\mathsf{L}} \)]is a \( 2\times N \) vector, where each row contains \( \sum _{\mathsf{L}}F^{+\left( \times \right) }_{\mathsf{L}}\mathbf{y}_{\mathsf{L}} \),
a vector combining data from all the detectors, weighted with the
inverse of the correlation matrix pertinent to each of them, and summed
with weights depending on the sky direction considered.
\item [\( \bm \Theta \equiv \sum _{\mathsf{L}}\mathsf{F}_{\mathsf{L}}\otimes \left( \mathbf{R}_{\mathsf{LL}}^{-1}\right) _{\parallel }\otimes \mathsf{F}_{\mathsf{L}}^{t} \)]is
a \( 2\times N_{\parallel }\times N_{\parallel }\times 2 \) matrix,
where as in the case of the single detector \( N_{\parallel } \)
is the dimension of the \( {\cal V}_{\parallel } \) subspace we are
testing for the presence of a burst. It is built summing the matrices
relative to each detector, then it must be inverted and contracted
with the \( 2\times N_{\parallel } \) matrices obtained restricting
\( \sum _{\mathsf{L}}\mathsf{F}_{\mathsf{L}}\otimes \mathbf{y}_{\mathsf{L}} \)
to \( {\cal V}_{\parallel } \) in order to construct the scalar statistic.
\end{description}
The matrix \( \mathsf{y} \) can be easily computed, while for \( \bm \Theta  \),
as in Eq.~\prettyref{eq:RinverseFromRmx}, the following identity
holds, in absence of signal:\begin{eqnarray}
E\left[ \mathsf{y}\otimes \mathsf{y}^{t}\right]  & = & \sum _{\mathsf{K},\mathsf{L}}\mathsf{F}_{\mathsf{K}}\otimes E\left[ \mathbf{y}_{\mathsf{K}}\otimes \mathbf{y}^{t}_{\mathsf{L}}\right] \otimes \mathsf{F}^{t}_{\mathsf{L}}\nonumber \\
 & = & \sum _{\mathsf{K},\mathsf{L}}\mathsf{F}_{\mathsf{K}}\otimes \mathbf{R}_{\left( \mathsf{KL}\right) y}\otimes \mathsf{F}^{t}_{\mathsf{L}}\, ;\label{eq:correlationYnetwork} 
\end{eqnarray}
if the detector noises are statistically independent, it reduces to\begin{equation}
E\left[ \mathsf{y}\otimes \mathsf{y}^{t}\right] =\sum _{\mathsf{L}}\mathsf{F}_{\mathsf{L}}\otimes \left( \mathbf{R}^{-1}_{\mathsf{LL}}\right) _{\parallel }\otimes \mathsf{F}^{t}_{\mathsf{L}}=\bm \Theta 
\end{equation}
where we have used \( \mathbf{R}_{(\mathsf{LL})y}=\mathbf{R}_{\mathsf{LL}}^{-1} \)
for the correlation matrix of the \( \mathbf{y}_{\mathsf{L}} \) time
series.

Now we would like to factor the correlation matrices \( \left( \mathbf{R}^{-1}_{\mathsf{LL}}\right) _{\parallel } \)
relative to each detector: however, each of them admits a different
KL expansion over the \( {\cal V}_{\parallel } \) subspace\begin{equation}
\left[ \mathbf{R}^{-1}_{\mathsf{LL}}\right] _{\parallel }=\sum _{k}\sigma ^{\mathsf{L}}_{k}\bm \psi _{\mathsf{L}}^{k}\otimes \bm \psi _{\mathsf{L}}^{k}
\end{equation}
and the bases \( \left\{ \bm \psi _{\mathsf{L}}^{k},\, k\in \left[ 1\, \, \ldots N_{\parallel }\right] \right\}  \)
are generally different for each detector%
\footnote{With the possible exception of detectors which are designed equal,
such as the two 4km LIGOs.
}, hence the sum of tensor products in Eq.~\prettyref{eq:correlationYnetwork}
does not factor in a patent way into a product of terms: however,
we are going to show that it can be factored introducing two DKL bases.

\subsubsection{\label{sec:generalNetworkCase}Exact form for the network statistic}

Let us fully exploit our understanding in Eq.~\prettyref{eq:correlationYnetwork}
that the matrix\begin{equation}
\label{eq:matrixTheta}
\bm \Theta =\sum _{\mathsf{L}}\left[ \begin{array}{cc}
\left( F^{+}_{\mathsf{L}}\right) ^{2} & F^{+}_{\mathsf{L}}F^{\times }_{\mathsf{L}}\\
F^{+}_{\mathsf{L}}F^{\times }_{\mathsf{L}} & \left( F^{\times }_{\mathsf{L}}\right) ^{2}
\end{array}\right] \otimes \left( \mathbf{R}^{-1}_{\mathsf{LL}}\right) _{\parallel }
\end{equation}
 is the correlation matrix of the (vector) signal \( \mathsf{y} \).
We note that the two time series \( \mathsf{y}_{+} \)and \( \mathsf{y}_{\times } \)
are \emph{jointly stationary}, that is also their cross correlation
depends just on the relative lag%
\footnote{Note also that the cross correlation is symmetric, because it is the
sum of symmetric matrices $\mathbf{R}^{-1}_\mathsf{K}$.
}. 

We can introduce two DKL bases, \( {\bm \psi _{+,\times }^{k},\, k\in \left[ 1,\, N_{\parallel }\right] } \),
for the two time series \( \mathsf{y}_{+,\times } \): in terms of
these two bases\begin{equation}
\label{eq:thetaMatrixExpanded}
\bm \Theta =\sum _{k,l}\, \left[ \begin{array}{cc}
\bm \sigma ^{++}_{kl}\bm \psi ^{k}_{+}\otimes \left[ \bm \psi ^{l}_{+}\right] ^{t}\,  & \bm \sigma ^{+\times }_{kl}\bm \psi ^{k}_{+}\otimes \left[ \bm \psi ^{l}_{\times }\right] ^{t}\\
\bm \sigma ^{\times +}_{kl}\bm \psi ^{k}_{\times }\otimes \left[ \bm \psi ^{l}_{+}\right] ^{t}\,  & \bm \sigma ^{\times \times }_{kl}\bm \psi ^{k}_{\times }\otimes \left[ \bm \psi ^{l}_{\times }\right] ^{t}
\end{array}\right] ;
\end{equation}
where the diagonal terms \( \left( \bm \sigma \right) ^{++} \) and
\( \left( \bm \sigma \right) ^{\times \times } \) are simple:\[
\left( \bm \sigma \right) _{kl}^{++\left( \times \times \right) }=\delta _{kl}\sigma ^{+\left( \times \right) }_{l},\]
where \( \sigma _{k}^{+},\sigma ^{\times }_{k} \)are the eigenvalues
of the two DKL bases, while the off diagonal terms are\begin{equation}
\label{eq:sigmaMatrix}
\bm \sigma ^{+\times \left( \times +\right) }_{kl}\equiv \sum _{\mathsf{K}}F_{\mathsf{K}}^{+}F_{\mathsf{K}}^{\times }\, \left[ \bm \psi ^{k}_{+\left( \times \right) }\right] ^{t}\cdot \left( \mathbf{R}^{-1}_{\mathsf{KK}}\right) _{\parallel }\cdot \bm \psi _{\times \left( +\right) }^{l}
\end{equation}
or simply in terms of the estimated cross correlations

\begin{eqnarray}
\bm \sigma ^{+\times \left( \times +\right) }_{kl} & \equiv  & \left[ \bm \psi ^{k}_{+\left( \times \right) }\right] ^{t}\cdot E\left[ y_{+\left( \times \right) }\otimes y_{\times \left( +\right) }\right] \cdot \bm \psi _{\times \left( +\right) }^{l}\nonumber \\
 & \equiv  & E\left[ c^{k}_{+\left( \times \right) }c_{\times \left( +\right) }^{l}\right] \label{eq:sigmaMatrix2} 
\end{eqnarray}
where \( c^{k}_{+\left( \times \right) } \) are the coefficients
of the DKL expansion in the two bases. Therefore the estimation of
the matrix \( \bm \sigma  \) is simple: once the \( \bm \psi _{+\left( \times \right) } \)
eigenvectors are defined, the eigenvalues give immediately the diagonal
terms \( \bm \sigma ^{++\left( \times \times \right) } \), while
the cross terms are most easily estimated from the data, performing
the DKL decomposition and cross correlating the coefficients.

The matrix \( \bm \Theta  \) can be easily inverted, obtaining \begin{equation}
\label{eq:thetaMatrixInverted}
\bm \Theta ^{-1}=\sum _{k,l}\left[ \begin{array}{cc}
\bm \rho ^{++}_{kl}\bm \psi ^{k}_{+}\otimes \left[ \bm \psi ^{l}_{+}\right] ^{t}\,  & \bm \rho ^{+\times }_{kl}\bm \psi ^{k}_{+}\otimes \left[ \bm \psi ^{l}_{\times }\right] ^{t}\\
\bm \rho ^{\times +}_{kl}\bm \psi ^{k}_{\times }\otimes \left[ \bm \psi ^{l}_{+}\right] ^{t}\,  & \bm \rho ^{\times \times }_{kl}\bm \psi ^{k}_{\times }\otimes \left[ \bm \psi ^{l}_{\times }\right] ^{t}
\end{array}\right] 
\end{equation}
 where \( \bm \rho \equiv \bm \sigma ^{-1} \) is such that\begin{equation}
\sum ^{N_{\parallel }}_{l=1}\sum _{q=+,\times }\bm \sigma _{kl}^{pq}\, \bm \rho _{lm}^{qr}=\delta _{km}\delta ^{pq}\, .
\end{equation}
It is easy to show that the solution for \( \bm \rho  \) is (note
that \( \bm \sigma ^{++\left( \times \times \right) } \) are diagonal
matrices)\begin{subequations}\begin{eqnarray}
\bm \rho ^{\times \times } & = & \left( \bm \sigma ^{\times \times }-\bm \sigma ^{\times +}\cdot \left( \bm \sigma ^{++}\right) ^{-1}\cdot \bm \sigma ^{+\times }\right) ^{-1}\label{eq:rhoMatrix} \\
\bm \rho ^{+\times } & = & -\left( \bm \sigma ^{++}\right) ^{-1}\cdot \bm \sigma ^{+\times }\cdot \bm \rho ^{\times \times }\label{eq:rhoMatrix2} 
\end{eqnarray}
\end{subequations}with similar expression exchanging \( \times ,+ \),
in terms of products of \( N_{\parallel }\times N_{\parallel } \)
matrices. We can finally write the statistic \( L=\mathsf{y}\cdot \left( \bm \Theta \right) ^{-1}\cdot \mathsf{y} \)
as a quadratic form over the coefficients of the two KL expansions:\begin{subequations}\begin{eqnarray}
L\!  & = & \! \sum ^{N_{\parallel }}_{k,l=1}\sum _{p,q}\bm \rho ^{pq}_{kl}\, c^{k}_{p}\, c^{l}_{q}\, \, \, ,\, \mathrm{where}\label{eq:LuncorrelatedNoise} \\
c^{k}_{p}\!  & \equiv  & \! \left[ \bm \psi ^{k}_{p}\right] ^{t}\! \cdot \left( \mathsf{y}_{p}\right) _{\parallel }=\! \sum ^{M}_{\mathsf{K}=1}\! F_{\mathsf{K}}^{p}\left[ \bm \psi ^{k}_{p}\right] ^{t}\! \cdot \left( \mathbf{R}^{-1}_{\mathsf{KK}}\cdot \mathbf{x}_{\mathsf{K}}\right) _{\parallel };\label{eq:cUncorrelatedNoise} 
\end{eqnarray}
\end{subequations}where we should keep in mind that the \( F^{\mathsf{I}}_{p} \)
functions depend the chosen direction in the sky, and with them both
the coefficients \( \sigma ^{pq}_{kl} \) and the DKL basis vectors
\( \bm \psi ^{k}_{p} \).

As argued in \prettyref{sec:exactStatistic}, the distribution(s)
of the DKL eigenvalues may allow to approximate the sum neglecting
terms which would result in noisy contributions: a in-depth discussion
would require however to consider a realistic noise spectrum.

We will come back to the algorithm defined by Eqs.~(\ref{eq:LuncorrelatedNoise},\ref{eq:cUncorrelatedNoise})
after having exposed the simplifications possible in the case of large
\( N_{\parallel } \).

\subsubsection{\label{sec:simplifiedNetworkCase}Simplified case: large \protect\( N_{\parallel }\protect \)}

As in \prettyref{sec:approximateStatistic}, if \( N_{\parallel } \)
is large enough to justify the approximation of the DKL with a Fourier
transform, the matrix \( \left( \mathbf{R}_{\left( \mathsf{KK}\right) y}\right) _{\parallel }=\left( \mathbf{R}^{-1}_{\mathsf{KK}}\right) _{\parallel } \)
can be written as follows:\begin{equation}
\left( \mathbf{R}_{\left( \mathsf{KK}\right) y}\right) _{\parallel }=\frac{f_{s}}{2}\sum ^{N_{\parallel }-1}_{k=0}S_{\left( \mathsf{KK}\right) y}\left[ k\right] \mathbf{w}_{k}\otimes \mathbf{w}^{H}_{k}
\end{equation}
where \( S_{\left( \mathsf{KK}\right) y}\left[ k\right]  \) is the
one-sided noise spectrum of the \( \left( \mathbf{y}_{\mathsf{K}}\right) _{\parallel } \)
data; its frequency resolution is \( f_{s}/N_{\parallel } \), and
it does not depend on the direction in the sky. Now the matrix \( \bm \Theta  \)
(see Eq.~\prettyref{eq:correlationYnetwork}) can be factored out
and we obtain in analogy with the results of Anderson \emph{et al}.\cite[Sec. V C]{anderson:brady:creighton:flanagan:01}\begin{equation}
\label{eq:thetaFattorizzata}
\sum _{\mathsf{K}}\mathsf{F}_{\mathsf{K}}\otimes \left( \mathbf{R}_{\mathsf{KK}}^{-1}\right) _{\parallel }\otimes \mathsf{F}^{t}_{\mathsf{K}}=\frac{f_{s}}{2}\sum ^{N_{\parallel }-1}_{k=0}\mathbf{w}_{k}\otimes \mathsf{S}_{y}\left[ k\right] \otimes \mathbf{w}^{H}_{k}
\end{equation}
where we have introduced a network spectral density \( \mathsf{S}_{\mathsf{y}} \)
for the \( \delta  \) filtered data \( \mathbf{y} \): each element
of \( \mathsf{S}_{\mathsf{y}} \) is a \( 2\times 2 \) matrix, depending
on the Euler angles \( \phi \, \theta  \) through the \( \mathsf{F}_{\mathsf{K}} \)
terms\begin{equation}
\label{eq:networkSpectralDensity}
\mathsf{S}_{\mathsf{y}}\left[ k\right] \equiv \sum _{\mathsf{K}}S_{\left( \mathsf{KK}\right) y}\left[ k\right] \, \mathsf{F}_{\mathsf{K}}\otimes \mathsf{F}^{t}_{\mathsf{K}}\, .
\end{equation}
We can therefore rewrite the statistic in Eq. \prettyref{eq:networkLogLikelyhood}
as\begin{subequations}\begin{eqnarray}
L\left( \mathsf{x}\right)  & = & \frac{2\, f_{s}}{N_{\parallel }}\sum ^{N_{\parallel }-2}_{k=1}\left[ \tilde{\mathsf{y}}\left[ k\right] \right] _{\parallel }^{H}\cdot \left[ \mathsf{S}_{\mathsf{y}}\left[ k\right] \right] ^{-1}\cdot \left[ \tilde{\mathsf{y}}\left[ k\right] \right] _{\parallel }\label{eq:networkStatisticSimplified} \\
\tilde{\mathsf{y}}\left[ \mathsf{k}\right]  & \equiv  & \sum ^{M}_{\mathsf{L}=1}\mathsf{F}_{\mathsf{L}}\tilde{y}_{\mathsf{L}}\left[ k\right] 
\end{eqnarray}
\end{subequations}in complete analogy with the single detector case,
Eq.~\prettyref{eq:statisticApproximate}; in this case \( \tilde{\mathsf{y}}\left[ \mathsf{k}\right]  \)
is a 2 elements vector, whose components are just the Fourier transform
of the \( \mathsf{y}_{+,\times } \) time series. In analogy with
the single detector case it is simple to prove that this statistic
converges to the excess power statistic defined in \cite[Eq. (5.29)]{anderson:brady:creighton:flanagan:01}
in the limit \( N_{\parallel }\rightarrow N \), or equivalently when
the cross correlations between the subspaces \( {\cal V}_{\parallel },{\cal V}_{\perp } \)
can be neglected.

\subsubsection{\label{sec:networkAlgorithm}Description of the algorithm}

We find useful to briefly outline the detection algorithm in the case
of uncorrelated noise across the detectors: we will see that the changes
induced by the correlations can be accommodated easily. We can be
sketchy because most steps are similar to those described in \prettyref{sec:implementation}.

Our \( M \)-detectors network produces \( M\times N \) data, represented
by the vectors \( \mathbf{x}_{\mathsf{K}} \); we are looking for
burst of length \( N_{\parallel }\ll N \). Then we should

\begin{enumerate}
\item filter the data of the individual detectors for the occurrence of
\( \delta  \) events, obtaining new vectors \( \mathbf{y}_{\mathsf{K}} \);
this operation can be done with the Fourier transform and costs \( O(M\times N\ln N) \).
\item For each direction in the sky, shift the \textbf{\( \mathbf{y}_{\mathsf{K}} \)}
vectors to compensate for the delays \( \tau _{I} \), and sum the
\( M \) vectors weighed with the \( F^{+\left( \times \right) }_{\mathsf{K}} \)
polarizations, to obtain \( y_{+,\times } \); cost \( N_{\mathrm{sky}}\times O(M\times N) \)
where we have introduced \( N_{\mathrm{sky}} \) as the number of
effectively independent directions.
\item Estimate the \( \bm \psi _{+,\times } \) DKL bases appropriate for
expanding \( \left( y_{+,\times }\right) _{\parallel } \). This step
may not be needed for large \( N_{\parallel } \); if required it
costs \( O\left( N_{\mathrm{sky}}\times N^{3}_{\parallel }\right)  \).
\item If using the DKL bases, build \( \bm \Theta ^{-1} \), that is compute
the \( 2\times N_{\parallel }\times N_{\parallel }\times 2 \) matrix
\( \left( \bm \sigma ^{-1}\right) _{lm}^{pq} \); this step requires
\( O(N_{\mathrm{sky}}\times N_{\parallel }^{3}) \) operations; otherwise,
estimate the {}``network'' spectral density \( \mathsf{S}_{y}\left[ k\right]  \)
of the \( y \) data, with frequency resolution \( f_{s}/N_{\parallel } \),
for instance averaging over the \( N/N_{\parallel } \) possible \( {\cal V}_{\parallel } \)
subspaces: a step costing \( O\left( \frac{N}{N_{\parallel }}\times N_{\parallel }\ln N_{\parallel }\right) +O\left( M\times N_{\parallel }\right)  \)
\item For each possible interval of length \( N_{\parallel } \) in the
data vector, perform the decomposition over the bases \( \bm \psi _{+,\times } \)
or alternatively over the Fourier basis: the cost is \( O\left( N_{\mathrm{sky}}\times N\times N_{\parallel }\right)  \)
in the first case, \( O\left( N_{\mathrm{sky}}\times N\times \ln N_{\parallel }\right)  \)
in the second case.
\item Evaluate the statistic \( L \) either using Eq.~\prettyref{eq:LuncorrelatedNoise}
or using Eq.~\prettyref{eq:networkStatisticSimplified}, depending
on the path followed. 
\end{enumerate}
In the large \( N_{\parallel } \) approximation several further optimizations
are possible: for example the Fourier transform of data \( \mathbf{y}_{\mathsf{L}} \)
can be done once, and the shifts needed to evaluate the sum \( \sum _{\mathsf{L}}\mathsf{F}_{\mathsf{L}}\otimes \tilde{\mathbf{y}}_{\mathsf{L}} \)
for different sky directions become phase factors to be attached to
the \( \mathsf{F}_{\mathsf{L}} \) tensors. The big unknown in this
estimation is the number \( N_{\mathrm{sky}} \) of independent sky
direction which should be probed. Given the {}``spin-2'' dependence
of the antenna patterns on the sky location, one expects a relatively
slow variation, hence it should be possible to sample the solid angle
on a reduced number of \( \phi ,\theta  \) angles, and possibly exploit
hierarchical methods to further reduce \( N_{\mathrm{sky}} \): this
may be the subject of future work.

\subsection{\label{sec:networkCorrelatedNoise}The case of correlated noise among
the detectors}

The problem of correctly writing down the likelihood in the case of
correlated noise among \( M \) detectors has already been studied
in depth by Finn \cite{finn:01}, who has also proposed to apply a
transformation to the \( M \) data channels, which would de-correlate
the noise. We face here a technical difficulty, such a transformation
would also {}``rotate'' the signal and render awkward the bookkeeping
in our derivation: we want here to explore a different approach, motived
by the hope that the cross-correlation terms will result to be significantly
smaller than the diagonal terms, so that a perturbation expansion
is possible.

If there is correlated noise, the likelihood for observing \( \bm x \)
in presence of a signal \( \mathsf{h} \) can be written in full generality
as \cite[Sec. III B]{finn:01}\begin{equation}
\Lambda \left( \bm x|\, \mathsf{h}\right) =\exp \left[ -\frac{1}{2}\left\langle \bm s,\bm s\right\rangle _{\bm R}+\left\langle \bm s,\, \bm x\right\rangle _{\bm R}\right] 
\end{equation}
where we have already defined the symbol \( \bm x \) for the \( M\times N \)
matrix representing the \( M \) time series, each of length \( N \),
produced by the detectors, while\begin{equation}
\bm s=\mathbf{s}_{1}\oplus \mathbf{s}_{2}\oplus \cdots \oplus \mathbf{s}_{M}
\end{equation}
is the signal at each detector, dependent on \( h_{+},\, h_{\times } \)
that we have defined in the wave frame. We borrow from\cite{finn:01}
the notation (with a different normalization)\begin{equation}
\left\langle \bm a,\, \bm b\right\rangle _{\bm R}\equiv \bm a\left( \bm R\right) ^{-1}\bm b=\sum _{\mathsf{KL}\, \mathsf{kl}}a_{K}\left[ k\right] \, \left( \bm R^{-1}\right) _{\mathsf{KL}\, kl}\, b_{L}\left[ l\right] ;
\end{equation}
\( \bm R \) is a \( M\, N\times M\, N \) matrix which we regard
as a tensor\begin{equation}
\bm R\equiv E\left[ \bm n\otimes \bm n\right] \, ,
\end{equation}
that is\begin{equation}
\left( \bm R\right) _{\mathsf{KL}\, kl}=\left( \mathbf{R}_{\mathsf{KL}}\right) _{kl}=E\left[ n_{\mathsf{K}}[k+d_{\mathsf{K}}]\, n_{\mathsf{L}}[l+d_{\mathsf{L}}]\right] \, .
\end{equation}

Each \( N\times N \) matrix \( \mathbf{R}_{\mathsf{KL}} \) is a
Toeplitz matrix depending only on \( k-l+\left( d_{\mathsf{K}}-d_{\mathsf{L}}\right)  \)
and not necessarily symmetric, unless \( \mathsf{K}=\mathsf{L} \);
only the symmetry \( \left( \bm R\right) _{\mathsf{KL}\, kl}=\left( \bm R\right) _{\mathsf{LK}\, lk} \)
holds, which ensures that \( \left\langle \bm a,\, \bm b\right\rangle =\left\langle \bm b,\, \bm a\right\rangle  \).
Notice also that, similarly to the previous sections, we have shifted
the labeling of data on each detector so that the burst is simultaneous
in the time series: we must be careful, because the time shifts do
not cancel out in the cross terms of the correlation matrix.

The inverse matrix \( \bm R^{-1} \)is defined, as in \cite{finn:01},
such that\begin{equation}
\delta _{\mathsf{IJ}}\delta _{ij}=\left( \bm R^{-1}\bm R\right) _{\mathsf{IJ}\, ij}=\sum ^{M}_{\mathsf{K}=1}\sum ^{N}_{k=1}\left( \bm R^{-1}\right) _{\mathsf{IK}\, ik}\left( \bm R\right) _{\mathsf{KJ}\, kj}
\end{equation}
and the network likelihood can be written explicitly as\begin{equation}
\Lambda \left( \bm x|\, \mathsf{h}\right) \equiv e^{-\frac{1}{2}s_{\mathsf{K}}\left[ k\right] \left( \bm R^{-1}\right) _{\mathsf{KL}\, kl}s_{\mathsf{L}}\left[ l\right] +s_{\mathsf{K}}\left[ k\right] \left( \bm R^{-1}\right) _{\mathsf{KL}\, kl}x_{\mathsf{L}}\left[ l+d_{\mathsf{L}}\right] };
\end{equation}
summation is implied over the indices \( k,l \) labeling the samples,
and the indices \( \mathsf{K},\mathsf{L} \) labeling the detectors.

\begin{widetext}

We have already written in Eq.~\prettyref{eq:signalVector} the explicit
form of \( s_{\mathsf{L}}\left[ l\right]  \), and we can proceed
as before to integrate over the nuisance parameters \( h_{+,\times } \):
we obtain formally a similar expression\begin{equation}
\label{eq:likelyhoodMultipleDetectors}
2\ln \Lambda \left( \bm x|\, \theta ,\phi \right) =\left[ \sum _{\mathsf{I}}\mathsf{F}_{\mathsf{I}}\otimes \mathbf{y}_{\mathsf{I}}\right] ^{t}_{\parallel }\cdot \left[ \sum _{\mathsf{KL}}\mathsf{F}_{\mathsf{K}}\otimes \left( \bm R^{-1}\right) _{\mathsf{KL}}\otimes \mathsf{F}^{t}_{\mathsf{L}}\right] _{\parallel }^{-1}\cdot \left[ \sum _{\mathsf{J}}\mathsf{F}_{\mathsf{J}}\otimes \mathbf{y}_{\mathsf{J}}\right] _{\parallel }=\mathsf{y}_{\parallel }\cdot \left[ \bm \Theta \right] _{\parallel }^{-1}\cdot \mathsf{y}_{\parallel }
\end{equation}
where again the suffix \( \parallel  \) means restriction of time
indices to the \( {\cal \bm V}_{\parallel } \) subspace; \( \bm \Theta  \)
is a \( 2\times N_{\parallel }\times N_{\parallel }\times 2 \) matrix,
constructed contracting the detector indices in \( \bm R \) with
the corresponding indices in the \( 2\times M \) matrix \( \mathsf{F} \),
while \( \mathsf{y}_{\parallel } \) is a \( 2\times N_{\parallel } \)
matrix obtained contracting the detector index of \( \mathsf{F} \)
with the corresponding index in the matrix \( \bm y \).

\end{widetext}

Notice that \( \mathbf{y}_{\mathsf{I}} \) combines data from the
different interferometers: one has by definition\begin{equation}
\label{eq:yShiftedInTime}
y_{\mathsf{I}}\left[ i\right] \equiv \sum _{\mathsf{J}\, j}\left( \bm R^{-1}\right) _{\mathsf{IJ}\, ij}x_{\mathsf{J}}\left[ j+d_{\mathsf{J}}\right] \, ,
\end{equation}
a sort of \( \delta  \) function filtering for multiple interferometers.

Given the formal exact solution in Eq.~\prettyref{eq:likelyhoodMultipleDetectors},
we proceed with the assumption that the cross correlations among detectors
are much smaller than the internal correlations: more precisely, that\begin{equation}
\frac{||\mathbf{R}_{\mathsf{KL}}||}{\sqrt{||\mathbf{R}_{\mathsf{KK}}||\, ||\mathbf{R}_{\mathsf{LL}}||}}\ll 1
\end{equation}
where \( ||\mathbf{A}||\equiv \max _{||\mathbf{x}||=1}||\mathbf{A}\cdot \mathbf{x}|| \)
is the matrix norm induced by the standard vector norm \( ||\mathbf{x}||\equiv \sqrt{\mathbf{x}\cdot \mathbf{x}} \).
We split \( \bm R \) into a block diagonal \( \bm D \), and a off-diagonal
\( \bm O \):\begin{subequations}\begin{eqnarray}
\bm R & = & \bm D+\bm O\\
\left( \bm D\right) _{\mathsf{KL}\, kl} & = & \delta _{\mathsf{KL}}\, \left( \mathbf{R}_{\mathsf{KK}}\right) _{kl}
\end{eqnarray}
 \end{subequations}and expand \( \bm R^{-1} \) in powers of \( \bm O \)
using the identity \begin{equation}
\left( \bm D+\bm O\right) ^{-1}=\bm D^{-1}-\bm D^{-1}\bm O\left( \bm D+\bm O\right) ^{-1}
\end{equation}
which allows to approximate

\begin{equation}
\bm R^{-1}\simeq \bm D^{-1}-\bm D^{-1}\bm O\left[ \bm D^{-1}-\bm D^{-1}\bm O\left( \bm D^{-1}-\ldots \right) \right] 
\end{equation}
to any desired precision. This expression is useful because the inverse
of the block diagonal matrix \( \bm D \) is simple

\begin{equation}
\left( \bm D^{-1}\right) _{\mathsf{KL}\, kl}=\delta _{\mathsf{KL}}\left( \mathbf{R}^{-1}_{\mathsf{KK}}\right) _{kl}\, ,
\end{equation}
in terms of inverse correlation matrices on each detector.

We can use this expansion to write down an approximate likelihood:
the \( \delta  \) filtered data \( \mathbf{y}_{\mathsf{K}} \) are

\begin{equation}
\mathbf{y}_{\mathsf{K}}\simeq \mathbf{R}^{-1}_{\mathsf{KK}}\cdot \mathbf{x}_{\mathsf{K}}-\sum _{\mathsf{L}\neq \mathsf{K}}\mathbf{R}^{-1}_{\mathsf{KK}}\cdot \mathbf{R}_{\mathsf{KL}}\cdot \mathbf{R}^{-1}_{\mathsf{LL}}\cdot \mathbf{x}_{\mathsf{L}}
\end{equation}
where we understand the shift of the data \( \mathbf{x} \), to simplify
the notation, and we keep only the first order in \( \mathbf{R}_{\mathsf{KL}} \).
This expression can be evaluated in the Fourier space\begin{equation}
\label{eq:yFirstOrder}
y_{\mathsf{K}}[l]\simeq \frac{2}{N}\sum ^{N-2}_{k=1}\frac{e^{-i2\pi kl/N}}{S_{\mathsf{KK}}\left[ k\right] }\left[ \tilde{x}_{\mathsf{K}}\left[ k\right] -\sum _{\mathsf{L}\neq \mathsf{K}}\frac{S_{\mathsf{KL}}\left[ k\right] }{S_{\mathsf{LL}}[k]}\tilde{x}_{\mathsf{L}}\left[ k\right] \right] \, ;
\end{equation}
and can be easily extended to higher orders, leading to\begin{equation}
\label{eq:yAllOrders}
\bm y=\frac{2}{N}\sum ^{N-2}_{k=1}e^{-i2\pi kl/N}\left[ \bm S\left[ k\right] \right] ^{-1}\cdot \bm x\left[ k\right] 
\end{equation}
where \( \bm x \) is the \( M\times N \) data {}``vector'' of
the network, and \( \bm S\left[ k\right]  \) is the \( M\times M \)
matrix with elements \begin{equation}
S_{\mathsf{KL}}\left[ k\right] \equiv \frac{2}{f_{s}}\mathbf{w}^{H}_{k}\cdot \mathbf{R}_{\mathsf{KL}}\cdot \mathbf{w}_{k}\, .
\end{equation}
To prevent misunderstandings we underline that this procedure is not
a whitening, and it does not correspond to defining uncorrelated data
channels: it is instead the multi-detector analogous of filtering
for the occurrence of \( \delta  \)-function events.

Having obtained in this way an approximation to some desired order
of the \( \mathsf{y}=\left( \begin{array}{c}
\mathsf{y}_{+}\\
\mathsf{y}_{\times }
\end{array}\right) =\sum _{\mathsf{J}}\mathsf{F}_{\mathsf{J}}\otimes \mathbf{y}_{\mathsf{J}} \) data matrix (of size \( 2\times N \)), we are actually able to proceed
with the same methods exploited in the uncorrelated noise case. In
order to test for the occurrence of a burst in the subspace \( {\cal V}_{\parallel } \),
we need to restrict the \( \mathsf{y} \) data to the burst subspace
\( {\cal V}_{\parallel } \) and define there the DKL bases \( \bm \psi _{+\times } \),~as
in \ref{sec:generalNetworkCase}, or Fourier bases if \( N_{\parallel } \)
is large enough.

We are then able to \emph{estimate} the matrix \( \bm \Theta =E\left[ \mathsf{y}\otimes \mathsf{y}\right]  \)
and compute the elements of the \( \bm \sigma  \) matrix (see Eq.~\prettyref{eq:thetaMatrixExpanded})
exactly as in that case: the diagonal elements \( \bm \sigma ^{++\left( \times \times \right) } \)
just as diagonal matrices built from the eigenvalues \( \sigma _{k}^{+\left( \times \right) } \)
, and the off diagonal matrices \( \bm \sigma ^{+\left( \times \right) } \)
using Eq.~\prettyref{eq:sigmaMatrix2}; all the remaining derivation
goes unchanged.

\subsubsection{\label{sec:simplifiedCorrelatedNoise}Simplified case: large \protect\( N_{\parallel }\protect \)}

As in \prettyref{sec:simplifiedNetworkCase}, a considerable simplification
is possible if the DKL \( \bm \psi _{+\left( \times \right) } \)
bases of the \( {\cal V}_{\parallel } \) space converge to the Fourier
bases: in this case we have, by the very definition \( \bm \Theta =E\left[ \mathsf{y}\otimes \mathsf{y}\right]  \),
and in analogy with Eq.~\prettyref{eq:thetaFattorizzata}\begin{subequations}\begin{eqnarray}
\sum _{\mathsf{IJ}}\mathsf{F}_{\mathsf{I}}\! \otimes \! \left( \bm R^{-1}\right) _{\mathsf{IJ}}\! \otimes \! \mathsf{F}^{t}_{\mathsf{J}}\!  & \equiv  & \! \frac{1}{2}\! \sum _{k}\! \mathbf{w}_{k}\! \otimes \! \mathsf{S}_{\mathsf{y}}\left[ k\right] \! \otimes \! \mathbf{w}^{H}_{k}\\
\mathsf{S}_{\mathsf{y}}\left[ k\right] \!  & \equiv  & \! \left( \begin{array}{cc}
S_{++}\left[ k\right]  & S_{+\times }\left[ k\right] \\
S_{\times +}\left[ k\right]  & S_{\times \times }\left[ k\right] 
\end{array}\right) \label{eq:networkSpectralDensityCN} 
\end{eqnarray}
\end{subequations}where in turn \( S_{pq}\left[ k\right]  \) are
the 4 possible cross-spectra, at frequency resolution \( f_{s}/N_{\parallel } \),
defined from the data \( y_{+},y_{\times } \). Hence the log-likelihood
has exactly the same expression as in the uncorrelated noise case\begin{subequations}\begin{eqnarray}
L\left( \mathsf{x}\right)  & = & \frac{2f_{s}}{N_{\parallel }}\sum ^{N_{\parallel }-2}_{k=1}\left[ \tilde{\mathsf{y}}\left[ k\right] \right] _{\parallel }^{H}\cdot \left[ \mathsf{S}_{\mathsf{y}}\left[ k\right] \right] ^{-1}\cdot \left[ \tilde{\mathsf{y}}\left[ k\right] \right] _{\parallel }\label{eq:networkStatisticCNSimplified} \\
\tilde{\mathsf{y}}\left[ \mathsf{k}\right]  & \equiv  & \sum ^{M}_{\mathsf{L}=1}\mathsf{F}_{\mathsf{L}}\, \tilde{y}_{\mathsf{L}}\left[ k\right] 
\end{eqnarray}
\end{subequations}with the difference that the \( y_{\mathsf{K}} \)
have been obtained combining data from the different detectors, in
a manner dependent on the direction in the sky; approximately as in
Eq.~\prettyref{eq:yFirstOrder}, or exactly as in Eq.~\prettyref{eq:yAllOrders}.

The algorithm described in \prettyref{sec:networkAlgorithm} goes
almost unchanged: the only real change is in the computation of the
data \( \mathbf{y}_{\mathsf{K}} \). In particular, we have to rearrange
steps (1) and (2) because now the \( \delta  \)-filtering must be
preceded by the time shift of the data \( \mathbf{x}_{\mathsf{L}} \):
the two operations no longer commute. All the other steps remain unaltered.

\subsection{\label{sec:networkStatisticalProperties}Distribution of the network
statistic}

Having given in \prettyref{sec:statisticalAnalysis} a demonstration
that the statistic \( L \) is a \( \chi ^{2} \) distributed variable,
we can generalize those results: we know that the network statistic\begin{equation}
L=\mathsf{y}_{\parallel }\cdot \bm \Theta ^{-1}\cdot \mathsf{y}_{\parallel }
\end{equation}
is written in terms of the inverse correlation matrix of the variables
\( \mathsf{y}_{\parallel } \) themselves, viz.\begin{equation}
\bm \Theta =E\left[ \mathsf{y}_{\parallel }\otimes \mathsf{y}_{\parallel }\right] 
\end{equation}
where the average is taken in absence of signal. Hence again \( L \)
is a \( \chi ^{2} \) variable, as can be directly checked considering
its first two moments\begin{subequations}\begin{eqnarray}
E\left[ L|H_{0}\right]  & = & E\left[ \mathsf{y}_{\parallel }\bm \Theta ^{-1}\mathsf{y}_{\parallel }\right] =\mathrm{tr}\left( \bm \Theta \bm \Theta ^{-1}\right) \nonumber \\
 & = & 2N_{\parallel }\\
E\left[ L^{2}|H_{0}\right]  & = & 2\, \mathrm{tr}\left( \bm \Theta \bm \Theta ^{-1}\bm \Theta \bm \Theta ^{-1}\right) +\left[ \mathrm{tr}\left( \bm \Theta \bm \Theta ^{-1}\right) \right] ^{2}\nonumber \\
 & = & 2N_{\parallel }\left( 2N_{\parallel }+2\right) 
\end{eqnarray}
\end{subequations}and generalizing with a bit of combinatorics to\begin{equation}
E\left[ L^{m}|H_{0}\right] =\prod ^{m}_{k=1}\left[ 2N_{\parallel }+2\left( k-1\right) \right] 
\end{equation}
which is the moment expansion of a \( \chi ^{2}\left( 2N_{\parallel }\right)  \)
variable. In the same way, in presence of a signal the distribution
\( d_{1}\left( L\right)  \) will be given by Eq.~\prettyref{eq:P1complete}
with the substitution \( N_{\parallel }\rightarrow 2N_{\parallel } \)
and the \textsf{SNR} defined as\begin{equation}
\label{eq:snrDefinitionNetwork}
\mathsf{SNR}=\frac{E\left[ L|H_{1}\right] -E\left[ L|H_{0}\right] }{\sqrt{2\left( 2N_{\parallel }\right) }}\, ;
\end{equation}
the extra factor of \( \sqrt{2} \) in the denominator results from
the definition of the signal in terms of two polarizations.

\subsection{\label{sec:networkDirectionalSensitivity}Example: network sensitivity
to \protect\( \delta \protect \) events}

The {}``network spectral density'' introduced in Eqs.~(\ref{eq:networkSpectralDensity},\ref{eq:networkSpectralDensityCN})
is a \( 2\times 2 \) matrix of spectra and (complex) cross-spectra,
which depends on the sky direction: it is therefore interesting to
derive some scalar quantity which can be plotted in a spherical projection
and give a visual idea of the sensitivity of the network.

As a simple example, motivated by the short duration of the impulsive
features in some of the model waveforms \cite[Fig. 2 (Model A)]{dimmelmeier:font:mueller:01},
we may consider the response of the network to a burst of duration
\( dt_{\mathrm{burst}}=1/f_{s} \), where \( f_{s} \) is the sampling
rate in the detectors, and having amplitudes \( A_{+},A_{\times } \)
in the two polarizations. We assume that the noise is uncorrelated
across the detectors, and that the data have been shifted to capture
the event in every data streams at the same time index \( a \): we
have \begin{equation}
\tilde{x}_{\mathsf{L}}\left[ k\right] =\left( A_{+}F^{+}_{\mathsf{L}}+A_{\times }F^{\times }_{L}\right) \frac{1}{f_{s}}e^{-j2\pi ka/N}
\end{equation}
 and the \( \delta  \)-filtered signal is (see Eq.~\prettyref{eq:inverseMatrixR})\begin{equation}
y_{\mathsf{L}}\left[ l\right] =\left( A_{+}F^{+}_{\mathsf{L}}+A_{\times }F^{\times }_{\mathsf{L}}\right) \frac{2}{f_{s}N}\sum ^{N-2}_{k=1}\frac{e^{i2\pi k\left( l-a\right) /N}}{S_{\mathsf{LL}}\left[ k\right] };
\end{equation}
projecting on the \( {\cal V}_{\parallel } \) subspace means setting
\( l=a \), hence\begin{equation}
y_{\mathsf{L}}\left[ a\right] =\left( A_{+}F^{+}_{\mathsf{L}}+A_{\times }F^{\times }_{\mathsf{L}}\right) \mathrm{rms}\left( y_{\mathsf{L}}\right) 
\end{equation}
where we have defined \begin{eqnarray}
\mathrm{rms}\left( y_{\mathsf{L}}\right)  & \equiv  & \frac{1}{f_{s}N}\sum ^{N/2-1}_{k=1}\frac{1}{S_{\mathsf{LL}}\left[ k\right] }\nonumber \\
 & \simeq  & \frac{1}{f^{2}_{s}}\int ^{f_{\mathsf{Nyquist}}}_{f_{\mathsf{seism}}}\frac{df}{S_{\mathsf{LL}}\left( f\right) };
\end{eqnarray}
this is the same quantity resulting from the analysis in \cite{bhawal:dhurandhar:95},
where burst signals with uniform spectrum in the detection band had
been considered. Next, we have\begin{equation}
\mathsf{y}_{\parallel }=\sum _{\mathsf{L}}\, \mathrm{rms}\left( y_{\mathsf{L}}\right) \left( A_{+}F^{+}_{\mathsf{L}}+A_{\times }F^{\times }_{\mathsf{L}}\right) \, \left( \begin{array}{c}
F^{+}_{\mathsf{L}}\\
F^{\times }_{\mathsf{L}}
\end{array}\right) 
\end{equation}
and, noticing that \( \left( \mathbf{R}^{-1}_{\mathsf{LL}}\right) [a,a]=\mathrm{rms}\left( \mathrm{y}_{\mathsf{L}}\right)  \)\begin{equation}
\bm \Theta =\sum _{\mathsf{L}}\mathrm{rms}\left( \mathrm{y}_{\mathsf{L}}\right) \left( \begin{array}{cc}
\left( F^{+}_{\mathsf{L}}\right) ^{2} & F^{+}_{\mathsf{L}}F^{\times }_{\mathsf{L}}\\
F^{+}_{\mathsf{L}}F^{\times }_{\mathsf{L}} & \left( F^{\times }_{\mathsf{L}}\right) ^{2}
\end{array}\right) \, ;
\end{equation}
the log-likelihood statistic \( L=\mathsf{y}_{\parallel }\cdot \bm \Theta ^{-1}\cdot \mathsf{y}_{\parallel } \)
can now be easily evaluated: we average over \( A_{+},A_{\times } \)
keeping their geometric mean \( A\equiv \sqrt{A^{2}_{+}+A^{2}_{\times }} \)
fixed, and evaluate the resulting \textsf{SNR} as in Eq.~\prettyref{eq:snrDefinitionNetwork}
for networks of interferometric detectors built out of different partitions
of the instruments currently under commissioning. Because of the \( f^{-2}_{s} \)
factor in \( \mathrm{rms}\left( \mathrm{y}_{\mathsf{L}}\right)  \),
the scale is set by the effective amplitude \( A\, dt=A\, f^{-1}_{s} \).

\begin{figure*}
{\centering \resizebox*{0.4\textwidth}{!}{\includegraphics{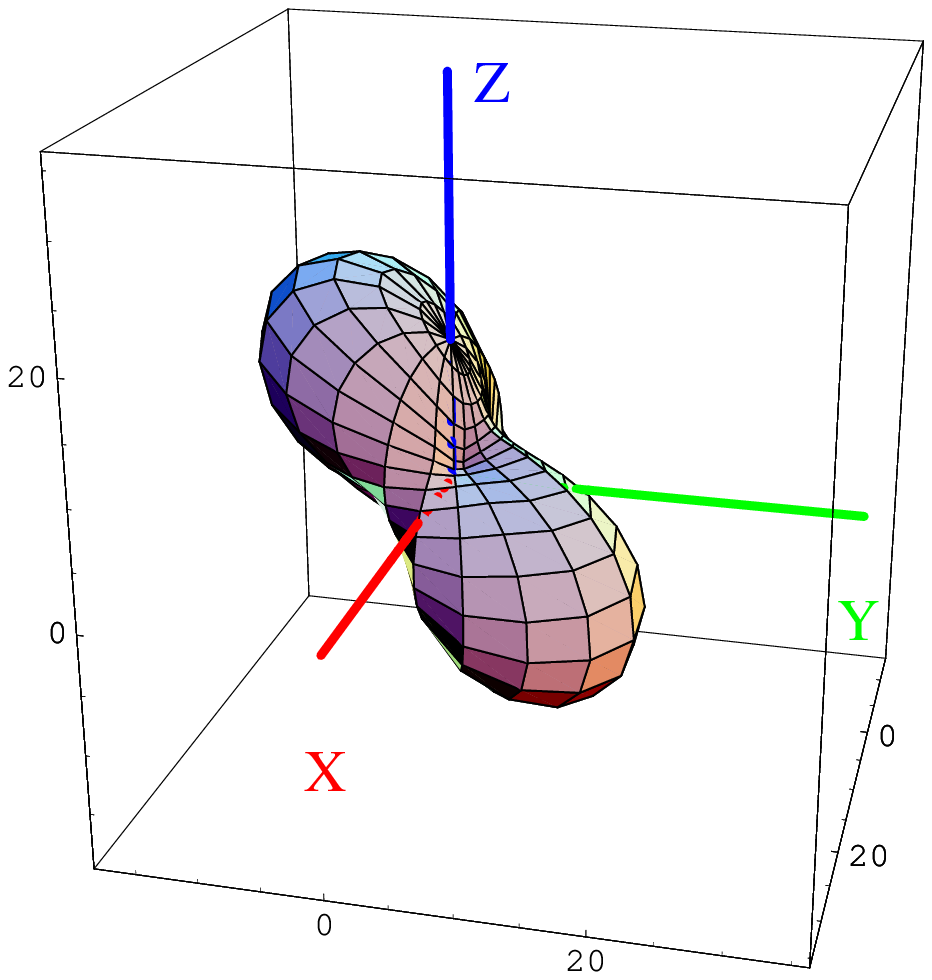}} \resizebox*{0.37\textwidth}{!}{\includegraphics{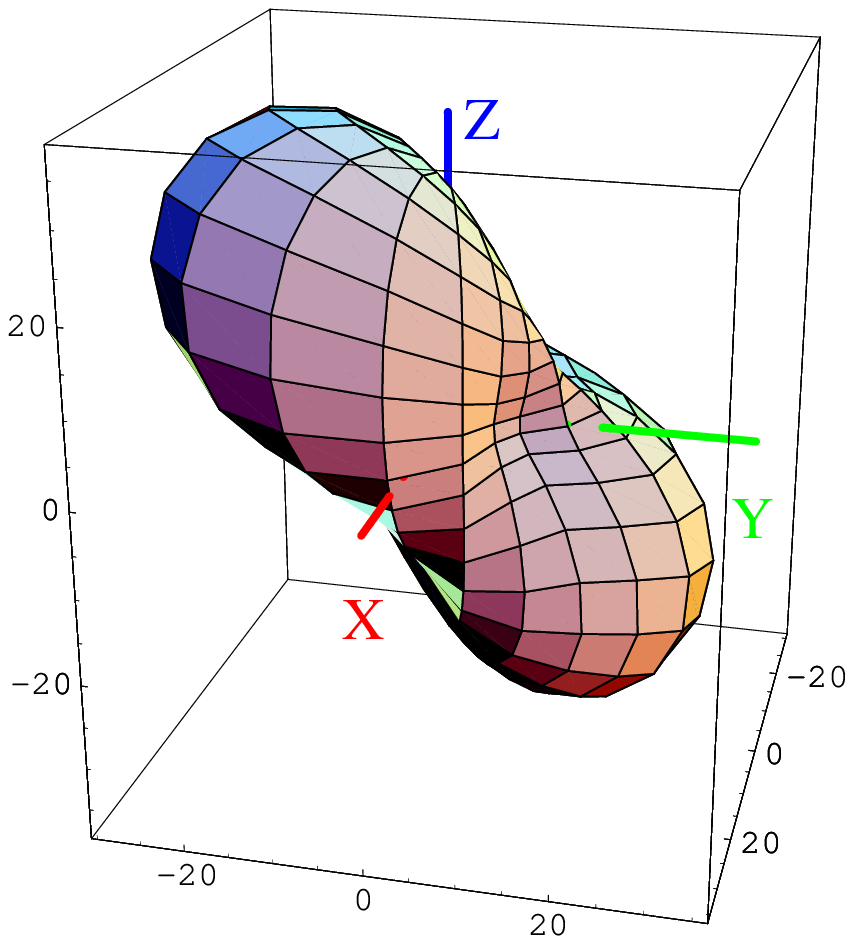}} \par}

\caption{\label{fig:sensitivityBurstLIGOandALL}Polar plot of the \textsf{SNR}
for bursts having \protect\( A\, dt=10^{-23}s\protect \), as a function
of the direction of the source. The figure at left refers to the network
of LIGO detectors, while the figure at right includes also GEO600,
TAMA and Virgo. The axes of the network frame are shown: we recall
that \textbf{Z} points toward the geographical north, and \textbf{X}
crosses the Greenwich meridian.}
\end{figure*}

We report in Fig.~(\ref{fig:sensitivityBurstLIGOandALL}) two polar
plots of the \textsf{SNR}, obtained setting \( A\, dt=10^{-23}s \);
with \( f^{-1}_{s}=O\left( 1ms\right)  \) this would correspond to
a strain \( A=O\left( 10^{-20}\right) , \) possible for a core collapse
event at a distance of 10 kpc \cite{zwerger:mueller:93,dimmelmeier:font:mueller:01}.
We have considered either the network of three LIGOs interferometers,
or a network including also GEO600, TAMA and Virgo: the details on
the detectors are reported in \prettyref{sec:networkCharacteristics},
where the nominal noise spectrum is modeled in Eq.~\prettyref{eq:noiseModel}
and in Tab.~\ref{tab:noiseParameters}, while locations and orientations
are reported in Tab.~\ref{tab:orientations}.

The plots have been obtained using a \emph{Mathematica\( ^{\mathrm{tm}} \)}
notebook which is available from the author upon request.

The global interferometric network appears significantly more sensitive,
and much of the effect is due to the contribution of Virgo: however
the result should be considered merely illustrative, because the chosen
shape of the burst (a \( \delta  \)-function) corresponds to a flat
spectrum in the frequency domain; this choice favors the Virgo detector
substantially, as already noticed in \cite{bhawal:dhurandhar:95},
because of the wide bandwidth of the model sensitivity of Virgo.

\section{\label{sec:conclusions}Conclusions and outlook}

In this paper we have defined a statistic for the detection of burst
signals, which is well suited to be applied to data affected by colored
noise, thus properly generalizing the \emph{excess power} statistic
\cite{flanagan:hughes:98,anderson:brady:creighton:flanagan:00,anderson:brady:creighton:flanagan:01}
to the case in which the spectral noise density varies significantly
over the frequency band of interest, and the signal prior is assumed
to be flat in \( \mathbb R^{N} \). It is optimal in the Bayes sense,
under the two hypotheses that the signal is distributed uniformly
in amplitude, and is contaminated by additive Gaussian noise. The
extension to the network case was straightforward and it was also
possible to take into account in a natural way the possible presence
of Gaussian noise correlations among the detectors, either perturbatively
or exactly.

The lack of assumptions on the GW signal distribution is both an advantage
and a disadvantage. We believe that the proposed statistic is the
correct one for a detection strategy free of \emph{a priori} assumptions,
apart the duration of the burst; yet we are aware that it does not
lend itself to easily include assumptions on the amplitude distribution
of the signal, as it was possible in \cite{anderson:brady:creighton:flanagan:01},
thus making difficult to set Bayesian thresholds.

Our generalization is not much more expensive, from the computational
point of view, than the statistic discussed by Anderson \emph{et al.}
\cite{anderson:brady:creighton:flanagan:00,anderson:brady:creighton:flanagan:01};
in its simplest implementation it amounts to perform a matched filtering
for \( \delta  \)-functions followed by the calculation of a {}``energy''
over the time window to be tested for the occurrence of a burst. The
whole analysis chain can be implemented combining a few standard time-series
analysis tools and therefore implemented with efficient algorithms.
The limited computational cost of the single detector search method
suggests that it may be applied to the full data set, before any triggering
is performed: in this sense it is a method proposed for the \emph{on-line}
search.

An evaluation of the actual detection performance, when considering
theoretical waveforms and simulated noise, remains to be done; this
will be the subject of future work, much along the lines of \cite{arnaud:99,pradier:00,arnaud:01}.

The cost of network detection was not fully estimated: it depends
on the number of directions in the sky effectively independent, which
was not studied in this paper.

Besides the detection of gravitational events, we propose this statistic
as a tool for excess noise characterization: real interferometric
detectors are definitely affected by non Gaussian noise \cite{allen-et-al:99},
and as long as the excess noise is dominated by burst signals, like
Poisson distributed creep events, we can think of using this statistic
as a tool for detecting and characterizing them. Once a event results
in a instance of the \( L \) statistic above the threshold, and therefore
a burst of excess noise or a gravitational wave is detected, our algorithm
provides a way to encode this information in a manner optimal with
respect to the distribution of the noise \cite{ortolan:vedovato:cerdonio:vitale:94}.
In fact, the discrete Karhunen-Lo\`eve transform that we have chosen
as a tool to compute the {}``energy'' over the burst time window
is equivalent to a principal component analysis: the coefficients
of the DKLT are ordered by the amount of RMS noise contributed by
the corresponding basis vectors. A candidate event can therefore be
encoded selecting just the largest DKLT coefficients, while retaining
most of its relevant {}``energy'', that is, the energy that is distributed
in less noisy components.

Moreover, we recall that in absence of signals the DKLT coefficients
are statistically uncorrelated: one may instead anticipate peculiar,
spurious correlations when a signal of any nature is present. These
correlations will emerge as regularities if the events occur repeatedly
in time, and it should be possible to catalog them in an automatic
way, for instance by using clustering methods in the vector space
of the DKLT coefficients.

\begin{acknowledgments}
The author gratefully thanks Alessandra Buonanno and Albert Lazzarini
for enlightnening discussions, and \'Eanna \'E. Flanagan and Jolien
D.E. Creighton for reading an earlier version of the manuscript and
providing me with valuable comments and suggestions. Warm thanks are
also due to the LSC reviewer, who spotted and corrected several misprints.

The LIGO Project and LIGO Laboratory are supported by the National
Science Foundation under cooperative agreement PHY-9210038.

This document has been assigned LIGO Document Control Center number
LIGO-P010019-01-E.
\end{acknowledgments}
\appendix

\section{\label{sec:conventions}DFT conventions}

We list here our conventions for the discrete time, discrete frequency
representation of stochastic processes. Let \( f_{s} \) be the sampling
frequency, and \( N \) the length of the data sample for the discrete
time process \( x\left[ l\right]  \); then the Fourier transform
pair \( \mathbf{x}\leftrightarrow \tilde{\mathbf{x}} \) is\begin{subequations}\begin{eqnarray}
\tilde{x}\left[ k\right]  & \equiv  & \frac{1}{f_{s}}\sum ^{N-1}_{l=0}e^{-i\, 2\pi \, k\, l/N}x[l],\\
x[l] & = & \frac{f_{s}}{N}\sum ^{N-1}_{k=0}e^{i\, 2\pi \, k\, l/N}\tilde{x}\left[ k\right] ;
\end{eqnarray}
\end{subequations}the one-sided sample spectrum \( S_{x}\left[ k\right]  \)
is \emph{defined} by\begin{equation}
\frac{1}{2}S_{x}\left[ k\right] =\frac{f_{s}}{N}E\left[ \left| \tilde{x}_{{\cal H}}\left[ k\right] \right| ^{2}\right] ;
\end{equation}
where \( \tilde{\mathbf{x}}_{{\cal H}}\leftrightarrow \left( {\cal H}*\mathbf{x}\right)  \)
is the Fourier transform of a suitably windowed realization of \textbf{\( \mathbf{x} \)}.
For large \( N \) one has that the correlation function \( R_{x}\left( \frac{a-b}{f_{s}}\right) =\left( \mathbf{R}_{x}\right) \left[ a,b\right]  \)
and the sample spectrum are \emph{approximately} Fourier pairs\begin{equation}
R_{x}\left( \tau \right) \simeq \frac{f_{s}}{N}\sum ^{N-1}_{k=0}\frac{1}{2}S_{x}\left[ k\right] \, e^{i\, 2\pi \, k\, \tau \, f_{s}/N};
\end{equation}
or equivalently\begin{equation}
\mathbf{R}_{x}\simeq f_{s}\sum ^{N-1}_{k=0}\frac{1}{2}S_{x}\left[ k\right] \, \mathbf{w}_{k}\otimes \mathbf{w}^{H}_{k},
\end{equation}
where \( \mathbf{w}_{k} \) are the Fourier orthonormal basis vectors
\begin{equation}
\mathbf{w}_{k}=\frac{1}{\sqrt{N}}\left[ 1,\omega ^{k},\omega ^{2k},\ldots \, \omega ^{(N-1)k}\right] 
\end{equation}
with \( \omega \equiv e^{i\, 2\pi /N} \). For a zero-mean process
one has\begin{equation}
\left( \mathbf{R}^{-1}_{x}\right) \left[ a,b\right] \simeq \frac{1}{f_{s}N}\sum ^{N-2}_{k=1}\frac{2}{S_{x}\left[ k\right] }e^{i\, 2\pi \left( a-b\right) k/N}
\end{equation}
and the useful relations\begin{subequations}\begin{eqnarray}
\left( \mathbf{R}_{x}\cdot \mathbf{y}\right) \left[ l\right]  & \simeq  & \frac{f^{2}_{s}}{N}\sum ^{N-1}_{k=0}\frac{1}{2}S_{x}\left[ k\right] \, \tilde{y}\left[ k\right] \, e^{i\, 2\pi kl/N}\\
\left( \mathbf{R}^{-1}_{x}\cdot \mathbf{y}\right) \left[ l\right]  & \simeq  & \frac{1}{N}\sum ^{N-2}_{k=1}\frac{2}{S_{x}\left[ k\right] }\tilde{y}\left[ k\right] \, e^{i\, 2\pi kl/N}
\end{eqnarray}
\end{subequations}where in the second one also \( y \) is a a zero
mean process. These are approximate relations because of the finite
\( N \).

\section{\label{sec:networkModel}Network model}

In this appendix we summarize the mathematics needed to describe a
network of detectors, the essential geometric characteristic of the
interferometers under construction, and their anticipated model spectral
densities.

\subsection{\label{sec:geometry}Geometry}

We adopt the following reference frames\cite{pai:dhurandhar:bose:01}:

\begin{description}
\item [\emph{network~frame}]centered on Earth and chosen with the \textbf{Z}
axis aligned along the geographical north, the \textbf{X} axis crossing
the Greenwich meridian;
\item [\emph{detector~frames}]centered on the beam splitter of each detector,
the \textbf{Z} axis pointing toward the local zenith and the \textbf{X}
axis bisecting the detector arms;
\item [\emph{wave~frame}]having the \textbf{Z} axis aligned along the direction
of propagation of the wave, and \textbf{X} axis lying in the \textbf{(X},
\textbf{Y)} plane of the network frame.
\end{description}
Rotations of coordinates from one frame to the other are expressed
in terms of Euler angles, that is\begin{subequations}\begin{eqnarray}
\mathsf{x}_{\mathsf{wave}} & = & \cal O(\phi ,\theta ,\psi )\cdot \mathsf{x}_{\mathsf{network}}\\
\mathsf{x}_{\mathsf{detector}_{\mathsf{L}}} & = & \cal O(\alpha _{\mathsf{L}},\beta _{\mathsf{L}},\gamma _{\mathsf{L}})\cdot \mathsf{x}_{\mathsf{network}}\, ;
\end{eqnarray}
\end{subequations}if \( \theta _{s},\phi _{s} \) are elevation and
azimuth of the source in the network frame, the relation with the
Euler angles \( \phi ,\theta  \) is\begin{equation}
\label{eq:polarAnglesVersusEulerAngles}
\phi =\phi _{s}-\frac{\pi }{2};\, \, \theta =\pi -\theta _{s}\, \, .
\end{equation}
The \( \psi  \) angle is zero, according to the definition of the
\emph{wave frame}. One can introduce the wave tensor\begin{equation}
\mathbf{w}(t)=\frac{1}{2}\left[ (h_{+}(t)+i\, h_{\times }(t))\, \mathbf{e}_{R}+(h_{+}(t)-i\, h_{\times }(t))\, \mathbf{e}_{L}\right] 
\end{equation}
where the helicity states \( \mathbf{e}_{R},\mathbf{e}_{L} \) can
be written as \begin{equation}
\mathbf{e}_{L,R}=\frac{1}{2}\left( \mathbf{e}_{X}\pm i\, \mathbf{e}_{Y}\right) \otimes \left( \mathbf{e}_{X}\pm i\, \mathbf{e}_{Y}\right) 
\end{equation}
in terms of unit vectors \( \mathbf{e}_{X},\mathbf{e}_{Y} \) specifying
the \textbf{X},\textbf{Y} axes of the wave frame as. In the network
frame they can also be written as Symmetric Trace Free tensors of
second rank (STF-2) \( {\cal Y}_{mn} \)\cite{gelfand:63,thorne:80}:\begin{eqnarray}
\mathbf{e}_{L,R} & = & \sqrt{\frac{8\pi }{15}}\left( {\cal Y}_{2\pm 2}\right) _{\mathsf{wave}}\nonumber \\
 & = & \sqrt{\frac{8\pi }{15}}T_{\pm 2n}\left( \phi ,\theta ,0\right) \left( {\cal Y}_{2n}\right) _{\mathsf{network}}
\end{eqnarray}
where \( T_{mn} \), with \( m,n=0,\pm 1,\pm 2 \) are Gel'fand's
functions of rank 2 and depend on the Euler angles \( \phi ,\theta ,\psi \left( =0\right)  \)
needed to rotate the network frame to the wave frame.

\begin{widetext}For completeness, the explicit form of the \( T^{l}_{mn} \)
functions for arbitrary rank is\cite{dhurandhar:tinto:88} as follows:\begin{eqnarray}
T^{l}_{mn}\left( \phi ,\theta ,\psi \right)  & = & e^{-i\left( n\phi +m\psi \right) }P^{l}_{mn}\left( \cos \theta \right) ,\\
P^{l}_{mn}\left( \mu \right)  & = & \frac{\left( -1\right) ^{l-m}i^{n-m}}{2^{l}\left( l-m\right) !}\sqrt{\frac{\left( l-m\right) !\left( l+n\right) !}{\left( l+m\right) !\left( l-n\right) !}}\frac{\left( 1-\mu \right) ^{\frac{m-n}{2}}}{\left( 1+\mu \right) ^{\frac{m+n}{2}}}\frac{d^{l-n}}{d\mu ^{l-n}}\left[ \frac{\left( 1+\mu \right) ^{m+l}}{\left( 1-\mu \right) ^{m-l}}\right] ,
\end{eqnarray}

where \( m,n\in \left[ -l,\, l\right]  \).\end{widetext}The detector
response is encoded by the tensor \( \mathbf{d}_{\mathsf{L}} \)\begin{equation}
\mathbf{d}_{\mathsf{L}}=\sin \left( 2\Omega _{\mathsf{L}}\right) \, \left( \mathbf{n}_{(\mathsf{L})1}\otimes \mathbf{n}_{(\mathsf{L})1}-\mathbf{n}_{(\mathsf{L})2}\otimes \mathbf{n}_{(\mathsf{L})2}\right) 
\end{equation}
in terms of the unit vectors aligned along the \( \mathsf{L} \)-th
interferometer arms, and the aperture angle \( 2\Omega _{\mathsf{L}} \)
of the arms. The factor \( \sin \left( 2\Omega _{\mathsf{L}}\right)  \)
is 1 for all the detectors apart \textsf{GEO600}, where it is \( 0.997 \);
consequently we will understand the factor in the following, to simplify
the notation.

The tensor \textbf{\( \mathbf{d} \)} has a simple expression\begin{equation}
\mathrm{tr}\left[ \mathbf{d}\cdot \left( {\cal Y}_{2m}\right) _{\mathsf{detector}}\right] =-i\sqrt{\frac{15}{8\pi }}\left( \delta _{m2}-\delta _{m-2}\right) 
\end{equation}
 in terms of STF-2 tensors in the detector frame. With two successive
rotations we obtain the coefficients of its expansion in terms of
STF-2 tensors in the wave frame\begin{subequations}\begin{eqnarray}
\mathrm{tr}\left[ \mathbf{d}\cdot \left( {\cal Y}_{2m}\right) _{\mathsf{network}}\right]  & = & -i\sqrt{\frac{15}{8\pi }}\Bigl [T^{\ast }_{2m}\left( \alpha _{\mathsf{L}},\beta _{\mathsf{L}},\gamma _{\mathsf{L}}\right) \nonumber \\
 &  & \qquad \quad -T^{\ast }_{-2m}\left( \alpha _{\mathsf{L}},\beta _{\mathsf{L}},\gamma _{\mathsf{L}}\right) \Bigr ]\\
\mathrm{tr}\left[ \mathbf{d}\cdot \left( {\cal Y}_{2m}\right) _{\mathsf{wave}}\right]  & = & \sum _{n}T_{mn}\left( \phi ,\theta ,0\right) \, \mathrm{tr}\left[ \mathbf{d}\cdot \left( {\cal Y}_{2n}\right) _{\mathsf{network}}\right] \nonumber \\
 & \equiv  & \sqrt{\frac{15}{8\pi }}D_{m}\left( \phi ,\theta ,0,\alpha _{\mathsf{L}},\beta _{\mathsf{L}},\gamma _{\mathsf{L}}\right) \label{eq:definitionD} 
\end{eqnarray}
\end{subequations}where we have introduced in the last formula a short-hand
notation.

Finally the signal at the \( \mathsf{L} \)-th detector, after proper
\( h \)-reconstruction in order to deconvolve the interferometer
response function, will be the scalar\begin{equation}
s_{\mathsf{L}}(t)=\mathrm{tr}\left[ \mathbf{w}\left( t-\tau _{\mathsf{L}}\left( \phi \, \theta \right) \right) \cdot \mathbf{d}_{\mathsf{L}}\right] 
\end{equation}
where \( \tau _{\mathsf{L}} \) is the delay at the \( \mathsf{L} \)-th
detector with respect to the network frame; it can be positive or
negative depending on the direction of the source. In terms of the
(complex) beam pattern functions \( F^{L,R}_{\mathsf{L}}\equiv \mathrm{tr}\left( \mathbf{e}_{L,R}\cdot \mathbf{d}_{\mathsf{L}}\right)  \)
for the two left and right wave polarizations\begin{subequations}\begin{eqnarray}
F^{L}_{\mathsf{L}} & = & D_{-2}\left( \phi ,\theta ,0,\alpha _{\mathsf{L}},\beta _{\mathsf{L}},\gamma _{\mathsf{L}}\right) \\
F^{R}_{\mathsf{L}}=\left( F^{L}_{\mathsf{L}}\right) ^{\ast } & = & D_{2}\left( \phi ,\theta ,0,\alpha _{\mathsf{L}},\beta _{\mathsf{L}},\gamma _{\mathsf{L}}\right) 
\end{eqnarray}
\end{subequations}one can alternatively write\begin{eqnarray}
s_{\mathsf{L}}(t) & = & \Re \left[ \left( h_{+}\left( t-\tau _{\mathsf{L}}\right) +i\, h_{\times }\left( t-\tau _{\mathsf{L}}\right) \right) \, F^{R}_{\mathsf{L}}\right] \\
 & = & \frac{1}{2}\left[ \left( h_{+}+i\, h_{\times }\right) D_{2}\left( \ldots \right) +\left( h_{+}-i\, h_{\times }\right) D_{-2}\left( \ldots \right) \right] \nonumber 
\end{eqnarray}
or equivalently and more conveniently for our work\begin{equation}
s_{\mathsf{L}}\left( t\right) =h_{+}F^{+}_{\mathsf{L}}+h_{\times }F^{\times }_{\mathsf{L}}
\end{equation}
where, reintroducing the aperture angle\begin{subequations}\begin{eqnarray}
F^{+}_{\mathsf{L}}\!  & \equiv  & \! \sin \left( 2\Omega _{\mathsf{L}}\right) \Re \left[ D_{-2}(\phi ,\theta ,0,\alpha _{\mathsf{L}},\beta _{\mathsf{L}},\gamma _{\mathsf{L}})\right] \\
F^{\times }_{\mathsf{L}}\!  & \equiv  & \! \sin \left( 2\Omega _{\mathsf{L}}\right) \Im \left[ D_{-2}(\phi ,\theta ,0,\alpha _{\mathsf{L}},\beta _{\mathsf{L}},\gamma _{\mathsf{L}})\right] \, .
\end{eqnarray}
\end{subequations}The given expression for the signal and for the
antenna patterns, as stressed in\cite{pai:dhurandhar:bose:01}, is
convenient because it keeps in factor form the rotations among the
various frames.

\subsection{\label{sec:networkCharacteristics}Interferometer network characteristics}

The geometrical characteristics of the detectors considered in this
study are listed in Table \ref{tab:orientations} where we quote the
latitude north of the equator, the longitude east of Greenwich, the
azimuths of the \textbf{X} and \textbf{Y} arms of the detectors, measured
counter-clockwise from the local east, and the \( \alpha ,\beta ,\gamma  \)
Euler angles needed to rotate coordinates in the network frame to
coordinates in the detector frame: all the angles are measured in
radians. The orientation data are taken from\cite{allen:96} and updated
with informations from the web sites of the different collaborations\cite{detector-locations:01};
the naming of the axes has been changed in one case so that all the
detectors have azimuth(\textbf{X} arm) < azimuth(\textbf{Y} arm).
The resulting Euler angles should be taken with care because are computed
in the approximation of spherical Earth and neglecting the elevation
of the detector sites and the fact that the arms are cords and not
tangents of the surface. For a more accurate model, please see \cite{accurateModel:01}.

\begin{table*}
\begin{tabular}{|c|c|c|c|c|c|}
\hline 
Detector&
Lat.&
Long.&
X azim.&
Y azim.&
\( \alpha ,\beta ,\gamma  \) Euler angles\\
\hline
\hline 
GEO600&
\( 0.911935 \)&
\( 0.171217 \)&
\( 0.377166 \)&
\( 2.02353 \)&
-1.20035, -0.658862, -1.74201\\
\hline 
LIGO Liv.&
\( 0.533373 \)&
\( -1.58424 \)&
\( 3.45575 \)&
\( 5.02655 \)&
2.04204, -1.03742, 0.013439\\
\hline 
LIGO Han.&
\( 0.810705 \)&
\( -2.0841 \)&
\( 2.21308 \)&
\( 3.78387 \)&
-2.99848, -0.760091, 0.513301\\
\hline 
TAMA&
\( 0.622733 \)&
\( 2.43543 \)&
\( 3.14159 \)&
\( 4.71239 \)&
2.35619, -0.948063, 2.27696\\
\hline 
Virgo&
\( 0.761487 \)&
\( 0.18326 \)&
\( 1.24791 \)&
\( 2.81871 \)&
-2.03331, -0.809309, -1.75406\\
\hline
\end{tabular}

\caption{\label{tab:orientations}detector locations and orientations, and
Euler angles (approximated) needed to express coordinates in the network
frame in terms of coordinates in the detector frames.}
\end{table*}

\begin{figure*}
{\centering \resizebox*{0.8\columnwidth}{!}{\includegraphics{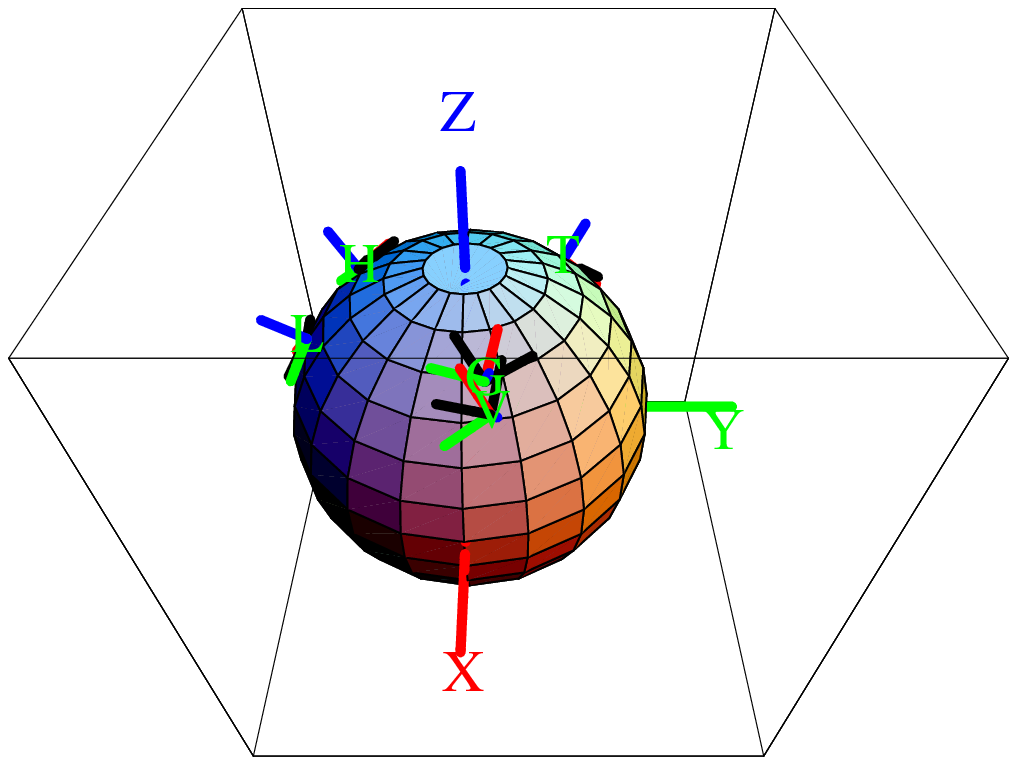}} \resizebox*{0.8\columnwidth}{!}{\includegraphics{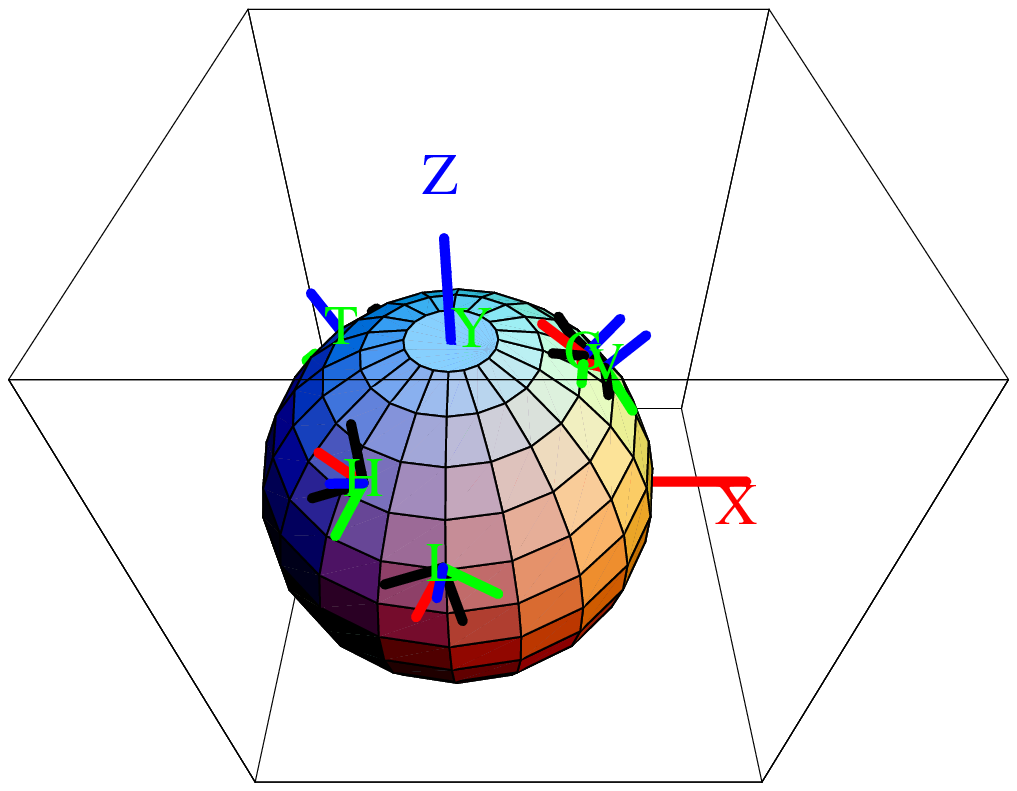}} \par}

\caption{\label{fig:detectorLocations}The locations of the detectors on Earth,
labeled by their initials (H and L for LIGO Hanford and LIGO Livingston);
the axes of the network frame, labeled \textbf{X},\textbf{Y},\textbf{Z}
and of the various detector frames are also shown. To the left: view
from above Europe. To the right: view from above the United States
of America.}
\end{figure*}

We show in Fig.~(\ref{fig:detectorLocations}) the locations of the
detectors and the reference frames attached to them, from two viewpoints
above Europe and the United States of America.

The other important characteristic of the detector is their planned
sensitivity. We have chosen to include only the baseline thermal and
shot-noise sources, omitting resonances in the observation band: the
model for the noise spectrum (filtered to deconvolve the response
function to gravitational waves) is therefore\begin{equation}
\label{eq:noiseModel}
S_{n}(f)=\frac{S_{\mathrm{pend}}}{f^{5}}+\frac{S_{\mathrm{mirror}}}{f}+S_{\mathrm{shot}}\left[ 1+\left( \frac{f}{f_{\mathrm{knee}}}\right) ^{2}\right] 
\end{equation}
where \( S_{\mathrm{pend}} \) quantifies the thermal noise of the
mirror pendular mode, above the pendulum resonance; \( S_{\mathrm{mirror}} \)
quantifies the \( 1/f \) tail of the internal modes of the mirror,
excited by thermal noise; \( S_{\mathrm{shot}} \) and \( f_{\mathrm{knee}} \)
parameterize the optical read-out noise.

In addition to these parameters, we call \( f_{\mathrm{seism}} \)
the cutoff below which the seismic noise is supposed to dominate over
the thermal noise. This simplified model does not include at least
two important effects, the thermal violin mode resonances and the
internal mirror resonance peaks, and should be considered merely illustrative.

\begin{figure*}
{\centering \resizebox*{0.8\textwidth}{!}{\includegraphics{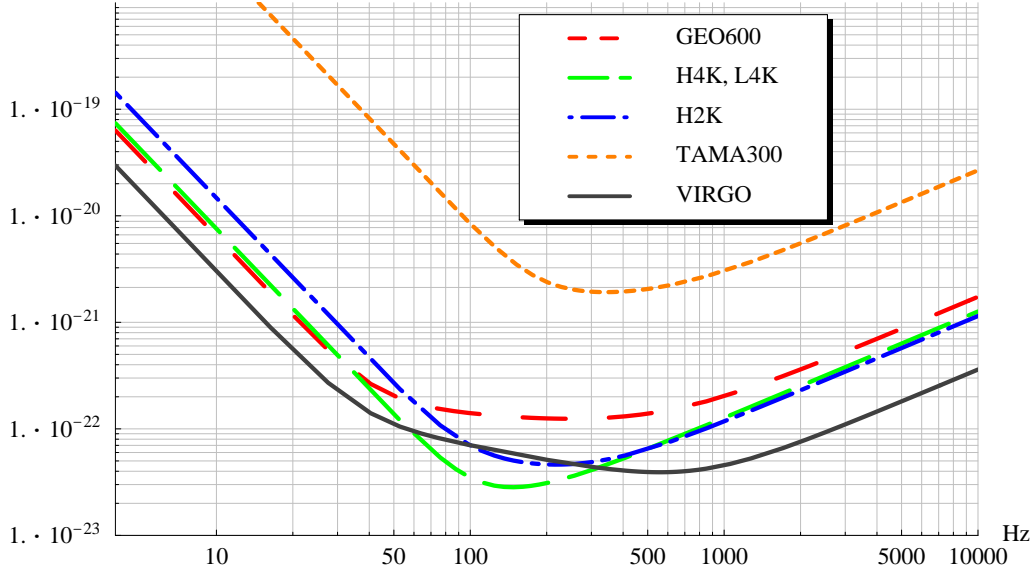}} \par}

\caption{\label{fig:noiseSpectra}A comparison of the baseline spectral sensitivities,
in units of Hz$^{-1/2}$, for the different interferometers currently
under commissioning, according to the simplified model in Eq.~\prettyref{eq:noiseModel}.
The two LIGO 4km interferometers {}``H4K'' and {}``L4K'', located
in Hanford and Livingston respectively, are supposed to have the same
sensitivity. For clarity's sake we do not display the so called {}``seismic
wall'' at \protect\( f_{\mathrm{seism}}\protect \), which should
be understood as dictated by Table \ref{tab:noiseParameters}.}
\end{figure*}

We report in Table~\ref{tab:noiseParameters} the numerical values
of these parameters, deduced from \cite{noise:spectra}, and in Fig.~(\ref{fig:noiseSpectra})
the comparison of the different noise spectral densities.

For the model at hand, the inverse correlation function\begin{equation}
R^{-1}_{n}(\tau )\propto \Re \int ^{\infty }_{0}e^{i2\pi f\tau }\frac{1}{S_{n}\left( f\right) }\, df
\end{equation}
 is well approximated by an expression of the form\begin{eqnarray}
R^{-1}_{n}\left( \tau \right)  & \simeq  & A_{0}e^{-\left| \tau \right| /\tau _{0}}\cos \left( 2\pi f_{0}\tau +\phi _{0}\right) \nonumber \\
 &  & +A_{1}e^{-\left| \tau \right| /\tau _{1}}\cos \left( 2\pi f_{1}\tau +\phi _{1}\right) \, ;\label{eq:RinverseApproximated} 
\end{eqnarray}
we list in Table~\ref{tab:ringingParameters} the values of the decay
times \( \tau _{0,1} \) and of the {}``ringing'' frequencies \( f_{0,1} \)
deduced from the values in Table~\ref{tab:noiseParameters}: the
small values of the \( \tau _{0,1} \) simply reflect the absence
of resonances in the model noise curves.

\begin{table}
\begin{tabular}{|c|c|c|c|c|c|}
\hline 
Detector&
\( f_{\mathrm{seism}} \)&
\( S_{\mathrm{pend}} \)&
\( S_{\mathrm{mirror}} \)&
\( S_{\mathrm{shot}} \)&
\( f_{\mathrm{knee}} \)\\
\hline
\hline 
GEO600&
\( 50 \)&
4.1e-36&
9e-43&
1e-44&
577\\
\hline 
LIGO 2K&
\( 40 \)&
2.1e-35&
2.25e-43&
4.35e-46&
182\\
\hline 
LIGO 4K&
\( 40 \)&
5.6e-36&
3.9e-44&
1.1e-46&
83\\
\hline 
TAMA&
\( 50 \)&
6.6e-31&
3.2e-40&
1.78e-42&
500\\
\hline 
Virgo&
\( 4 \)&
9e-37&
4.5e-43&
3.24e-46&
500\\
\hline
\end{tabular}

\caption{\label{tab:noiseParameters}Parameters characterizing the baseline
noise of the detectors in the network; cfr. Eq.~\prettyref{eq:noiseModel}.}
\end{table}

\begin{table}
\begin{tabular}{|c|c|c|c|c|}
\hline 
Detector&
\( \tau _{0} \)&
\( f_{0} \)&
\( \tau _{1} \)&
\( f_{1} \)\\
\hline
\hline 
GEO600&
5.6e-3&
32&
2.7e-4&
44\\
\hline 
LIGO 2K&
2.4e-3&
70&
6.1e-4&
107\\
\hline 
LIGO 4K&
2.5e-3&
83&
1.1e-3&
51\\
\hline 
TAMA&
1.4e-3&
141&
3.1e-4&
82\\
\hline 
Virgo&
6.0e-3&
27&
2.2e-4&
293\\
\hline
\end{tabular}

\caption{\label{tab:ringingParameters}The characteristic ringing frequencies
(measured in Hz) and correlation times (in sec.), for each interferometer,
deduced from the simplified noise models we have adopted.}
\end{table}


\begin{thebibliography}{10}
\bibitem{LIGO:92}A.Abramovici \emph{et al.} Science \textbf{256}, 325 (1992).
\bibitem{VIRGO:97}B.Caron \emph{et al.} Nucl. Phys. B (Proc.Suppl.) \textbf{54}, 167
(1997).
\bibitem{GEO:95}K.Danzmann \emph{et al.}, in \emph{First Edoardo Amaldi Conference
on Gravitational wave experiments,} ed. E.Coccia \emph{et al.} (World
Scientific, Singapore, 1995).
\bibitem{TAMA}K.Tsubono, in \emph{First Edoardo Amaldi Conference on Gravitational
Wave Experiments}, ed. E.Coccia \emph{et al.} (World Scientific, Singapore,
1995).
\bibitem{owen:sathya:99}B.J.Owen and B.S.Sathyaprakash, Phys. Rev. D \textbf{60}, 022002 (1999).
\bibitem{pradier:01}T.Pradier, N.Arnaud, M.A.Bizouard, F.Cavalier, M.Davier and P.Hello,
Phys. Rev. D \textbf{63,} 042002 (2001).
\bibitem{pradier:00}T.Pradier, N.Arnaud, M.A.Bizouard, F.Cavalier, M.Davier and P.Hello,
Int. J. Mod. Phys. D \textbf{9}, 309-314 (2000).
\bibitem{arnaud:01}N.Arnaud, M.Barsuglia, M.-A.Bizouard, P.Canitrot, F.Cavalier, M.Davier,
P.Hello and T.Pradier, \emph{Detection in coincidence of gravitational
wave bursts with a network of interferometric detectors (I): Geometric
acceptance and timing}, gr-qc/0107081.
\bibitem{zwerger:mueller:93}T.Zwerger and E.Müller, Astron. Astrophys. \textbf{267}, 623 (1993).
\bibitem{arnaud:99}N. Arnaud, F. Cavalier, M. Davier and P.Hello, Phys. Rev. D \textbf{59},
082002 (1999).
\bibitem{flanagan:hughes:98}\'E.\'E.Flanagan, S.A.Hughes, Phys. Rev. D \textbf{57}, 4566 (1998).
\bibitem{anderson:brady:creighton:flanagan:00}W.G.Anderson, P.R.Brady, J.D.E.Creighton, \'E.\'E.Flanagan, Int. J.
Mod. Phys. D \textbf{9}, 303 (2000).
\bibitem{anderson:brady:creighton:flanagan:01}W.G.Anderson, P.R.Brady, J.D.E.Creighton, \'E.\'E.Flanagan, Phys.
Rev. D \textbf{63}, 042003 (2001).
\bibitem{dimmelmeier:font:mueller:01}H.Dimmelmeier, J.A.Font and E.Müller, \emph{Gravitational waves from
relativistic core collapse,} astro-ph/0103088 (2001).
\bibitem{pizzella:93}G. Pizzella, \emph{Fisica sperimentale del campo gravitazionale},
(La Nuova Italia, Firenze, 1993).
\bibitem{papoulis:91}A.Papoulis, \emph{Probability, random variables, and stochastic processes}
(McGraw-Hill, New York, 1991).
\bibitem{therrien:92}C.W.Therrien, \emph{Discrete random signals and statistical signal
processing} (Prentice-Hall, Englewood Cliffs, 1992).
\bibitem{ortolan:vedovato:cerdonio:vitale:94}A.Ortolan, G.Vedovato, M.Cerdonio and S.Vitale, Phys. Rev. D \textbf{50},
4737 (1994).
\bibitem{cella:chickarmane:vicere:00}G.Cella, V.Chickarmane and A.Vicer\'e, Int. J. Mod. Phys. D \textbf{9},
269 (2000).
\bibitem{finn:01}L.S.Finn, Phys. Rev. D \textbf{63}, 102001 (2001).
\bibitem{donough:whalen:95}R.N.Donough and A.D.Whalen, \emph{Detection of Signals in Noise} (Academic
Press, San Diego, 1995).
\bibitem{ligo40mData:96}Abramovici \emph{et al.}, Phys. Lett. A \textbf{218}, 157-163 (1996).
\bibitem{finn:mukherjee:01}L.S.Finn and S.Mukherjee, Phys. Rev. D \textbf{63}, 062004 (2001).
\bibitem{schutz:91}B.F.Schutz, \emph{Data processing, analysis and storage for interferometric
antennas}, in \emph{The detection of gravitational waves}, ed. D.G.Blair,
(Cambridge Univ. Press, Cambridge, 1991).
\bibitem{cuoco:01}E.Cuoco, G.Calamai, L.Fabbroni, G.Losurdo, M.Mazzoni, R.Stanga and
F.Vetrano, Class. Quant. Grav. \textbf{18}, 1727-1752 (2001).
\bibitem{cuoco:02}E.Cuoco, G.Losurdo, G.Calamai, L.Fabbroni, M.Mazzoni, R.Stanga, G.Guidi
and F.Vetrano, \emph{Noise parametric identification and whitening
for LIGO 40-meter interferometer data,} gr-qc/0104071 (2001).
\bibitem{finn:92}L.S.Finn, Phys. Rev. D \textbf{46}, 5236 (1992).
\bibitem{percival:walden:93}D.B.Percival and A.T.Walden, \emph{Spectral analysis for physical
applications}, (Cambridge Univ. Press, Cambridge, 1993).
\bibitem{sintes:schutz:98}A.M.Sintes, B.F.Schutz, Phys. Rev. D \textbf{58}, 122003 (1998).
\bibitem{flaminio:00}D.Buskulic, L.Derome, R.Flaminio, R.Hermel, J.C.Lacotte, F.Marion,
A.Masserot, L.Massonnet, B.Mours, R.Morand, P.Puppo, D.Verkindt and
M.Yvert, Int. J. of Mod. Phys D \textbf{9}, 263-267 (2000).
\bibitem{helstrom:60}C.W.Helstrom, \emph{Statistical Theory of Signal Detection} (Pergamon
Press, New York, 1960).
\bibitem{gradshteyn:ryzhik:65}I.S.Gradshteyn and I.M.Ryzhik, \emph{Table of Integrals, Series, and
Products} (Academic Press, New York, 1965).
\bibitem{allen-et-al:99}B.Allen \emph{et al.}, Phys. Rev. Lett. \textbf{83}, 1498-1501 (1999).
\bibitem{dhurandhar:tinto:88}S.Dhurandhar and M.Tinto, Mon. Not. R. astr. Soc. \textbf{234}, 663-676
(1988).
\bibitem{gelfand:63}M.Gel'fand, R.A.Minlos and Z.Ye.Shapiro, \emph{Representations of
the Rotation and Lorentz Groups and their Applications} (Pergamon
Press, New York, 1963)\emph{.}
\bibitem{thorne:80}K.S.Thorne, Rev. Mod. Phys. \textbf{52}, 299-339 (1980).
\bibitem{pai:dhurandhar:bose:01}A.Pai, S.Dhurandhar and S.Bose, Phys. Rev. \textbf{}D \textbf{64},
042004 (2001).
\bibitem{lazzarini:vicere:01}A.Lazzarini and A.Vicer\'e, \emph{Analysis of the effects of long-term
correlations that may compromise long-integration cross-correlation
measurements}, LIGO technical note LIGO-T010101-00-E (2001).
\bibitem{christensen:90}N.Christensen, Ph.D. thesis, Massachussetts Institute of Technology,
(1990).
\bibitem{bhawal:dhurandhar:95}B.Bhawal and S.V.Dhurandhar, \emph{Coincidence detection of broadband
signals by networks of the planned interferometric gravitational wave
detectors}, (Ganeshkhind, IUCAA), IUCAA-29-95, gr-qc/9509042 (1995).
\bibitem{allen:96}B.Allen, \emph{Gravitational wave detector sites}, gr-qc/9607075 (1996).
\bibitem{detector-locations:01}Sources of information about detector locations and orientations:\\
\href{http://www.geo600.uni-hannover.de/geo600/project/location.html}{www.geo600.uni-hannover.de/geo600/project/location.html}\\
\href{http://www.ligo.caltech.edu/docs/T/T980044-08.pdf}{www.ligo.caltech.edu/docs/T/T980044-08.pdf}\\
\href{http://tamago.mtk.nao.ac.jp/tama/facility/faciliti.html}{tamago.mtk.nao.ac.jp/tama/facility/faciliti.htm}\\
\href{http://wwwlapp.in2p3.fr/virgo/virgo.html}{wwwlapp.in2p3.fr/virgo/virgo.html}
\bibitem{noise:spectra}Reference sites for the noise curves cited in the text: \\
\href{http://www.geo600.uni-hannover.de/geo600/project/wideband.html}{www.geo600.uni-hannover.de/geo600/project/wideband.html}\\
\href{http://www.ligo.caltech.edu/~kent/ASIS_NM/noise_models.html}{www.ligo.caltech.edu/\~{}kent/ASIS\_NM/noise\_models.html}
\\
\href{http://tamago.mtk.nao.ac.jp/tama/recom/recom3/index.html}{tamago.mtk.nao.ac.jp/tama/recom/recom3/index.html}\\
\href{http://www.virgo.infn.it/senscurve}{www.virgo.infn.it/senscurve}
\bibitem{accurateModel:01}W.G.Anderson, P.R.Brady, D.Chin, J.D.E.Creighton, K.Riles and J.T.Whelan,
\emph{Beam Pattern Response Functions and Times of Arrival for Earthbound
Interferometers}, \href{http://phys.utb.edu/UTBRG/activities/papers}{http://phys.utb.edu/UTBRG/activities/papers}\end{thebibliography}
\end{document}